\documentclass[aps,prb,10pt,onecolumn,superscriptaddress,nobibnotes]{revtex4-1}
\pdfoutput=1

\usepackage{lmodern}
\usepackage{ifpdf}
\usepackage[T1]{fontenc}
\usepackage[utf8]{inputenc}
\usepackage{graphicx}
\usepackage{amsmath, amsfonts, amssymb}
\usepackage[unicode=true,
            colorlinks=true,
            linkcolor=blue,
            citecolor=blue,
            urlcolor=blue]{hyperref} 
\usepackage{siunitx}
\usepackage{multirow}
\usepackage{upgreek}

\usepackage{colortbl}
\usepackage{footnote}
\usepackage{amssymb}			
\usepackage{hyperref}		
\usepackage[flushleft]{threeparttable}

\usepackage{float}

\usepackage[font=small,labelfont=bf,
justification=justified,
format=plain]{caption} 

\usepackage{subfig}

\usepackage{footnote}
\usepackage{booktabs,caption}
\usepackage[flushleft]{threeparttable}
\usepackage[bottom]{footmisc}				

\usepackage{ulem}

\usepackage[x11names]{xcolor}

\newcommand{\domark}{%
  \vbox to 0pt{
    \kern-\dp\strutbox
    \smash{\llap{\color{red!90!black}\#\kern0.5em}}
    \vss
  }%
}

\immediate\write18{git rev-parse --short HEAD > \jobname.gitinfo }
\newcommand{\gitrev}{\InputIfFileExists{\jobname.gitinfo}{}{ref not available}}

\begin{document}

\title{Electrical Spectroscopy of Polaritonic Nanoresonators}

\author{Sebasti\'{a}n Castilla} \email{sebastian.castilla@icfo.eu} \thanks{These authors contributed equally} \affiliation{ICFO - Institut de Ci\`{e}ncies Fot\`{o}niques, The Barcelona Institute of Science and Technology, Castelldefels (Barcelona) 08860, Spain}
\author{Hitesh Agarwal} \thanks{These authors contributed equally} \affiliation{ICFO - Institut de Ci\`{e}ncies Fot\`{o}niques, The Barcelona Institute of Science and Technology, Castelldefels (Barcelona) 08860, Spain}
\author{Ioannis Vangelidis} \thanks{These authors contributed equally} \affiliation{Department of Materials Science and Engineering, University of Ioannina, 45110 Ioannina, Greece}
\author{Yuliy Bludov} \thanks{These authors contributed equally} \affiliation{Centro de Física das Universidades do Minho e do Porto (CF-UM-UP) e Departamento de Física, Universidade do Minho, P-4710-057 Braga, Portugal}
\author{David Alcaraz Iranzo} \affiliation{ICFO - Institut de Ci\`{e}ncies Fot\`{o}niques, The Barcelona Institute of Science and Technology, Castelldefels (Barcelona) 08860, Spain}
\author{Adrià Grabulosa} \affiliation{ICFO - Institut de Ci\`{e}ncies Fot\`{o}niques, The Barcelona Institute of Science and Technology, Castelldefels (Barcelona) 08860, Spain}
\author{Matteo Ceccanti} \affiliation{ICFO - Institut de Ci\`{e}ncies Fot\`{o}niques, The Barcelona Institute of Science and Technology, Castelldefels (Barcelona) 08860, Spain}
\author{Mikhail I. Vasilevskiy} \affiliation{Centro de Física das Universidades do Minho e do Porto (CF-UM-UP) e Departamento de Física, Universidade do Minho, P-4710-057 Braga, Portugal} \affiliation{International Iberian Nanotechnology Laboratory (INL), Av Mestre José Veiga, 4715-330 Braga, Portugal}
\author{Roshan Krishna Kumar} \affiliation{ICFO - Institut de Ci\`{e}ncies Fot\`{o}niques, The Barcelona Institute of Science and Technology, Castelldefels (Barcelona) 08860, Spain}
\author{Eli Janzen} \affiliation{Tim Taylor Department of Chemical Engineering, Kansas State University, Manhattan, KS, USA}
\author{James H. Edgar} \affiliation{Tim Taylor Department of Chemical Engineering, Kansas State University, Manhattan, KS, USA}
\author{Kenji Watanabe} \affiliation{Research Center for Electronic and Optical Materials, National Institute for Materials Science, 1-1 Namiki, Tsukuba 305-0044, Japan}
\author{Takashi Taniguchi} \affiliation{Research Center for Materials Nanoarchitectonics, National Institute for Materials Science,  1-1 Namiki, Tsukuba 305-0044, Japan}
\author{Nuno Peres} \affiliation{Centro de Física das Universidades do Minho e do Porto (CF-UM-UP) e Departamento de Física, Universidade do Minho, P-4710-057 Braga, Portugal} \affiliation{International Iberian Nanotechnology Laboratory (INL), Av Mestre José Veiga, 4715-330 Braga, Portugal} \affiliation{POLIMA—Center for Polariton-driven Light-Matter Interactions, University of Southern Denmark, Campusvej 55, DK-5230 Odense M, Denmark}
\author{Elefterios Lidorikis} \affiliation{Department of Materials Science and Engineering, University of Ioannina, 45110 Ioannina, Greece} \affiliation{University Research Center of Ioannina (URCI), Institute of Materials Science and Computing, 45110 Ioannina, Greece}
\author{Frank H.L. Koppens} \email{frank.koppens@icfo.eu} \affiliation{ICFO - Institut de Ci\`{e}ncies Fot\`{o}niques, The Barcelona Institute of Science and Technology, Castelldefels (Barcelona) 08860, Spain} \affiliation{ICREA - Instituci\'o Catalana de Recerca i Estudis Avan\c{c}ats, 08010 Barcelona, Spain}

\maketitle
\section*{\textsf{Abstract}}
{\bf
One of the most captivating properties of polaritons is their capacity to confine light at the nanoscale. This confinement is even more extreme in two-dimensional (2D) materials. 2D polaritons have been investigated by optical measurements using an external photodetector. However, their effective spectrally resolved electrical detection via far-field excitation remains unexplored. This fact hinders their potential exploitation in crucial applications such as sensing molecules and gases, hyperspectral imaging and optical spectrometry, banking on their potential for integration with silicon technologies. Herein, we present the first electrical spectroscopy of polaritonic nanoresonators based on a high-quality 2D-material heterostructure, which serves at the same time as the photodetector and the polaritonic platform. We employ metallic nanorods to create hybrid nanoresonators within the hybrid plasmon-phonon polaritonic medium in the mid and long-wave infrared ranges. Subsequently, we electrically detect these resonators by near-field coupling to a graphene pn-junction. The nanoresonators simultaneously present a record of lateral confinement and high-quality factors of up to $\sim$200, exhibiting prominent peaks in the photocurrent spectrum, particularly at the underexplored lower reststrahlen band of hBN. We exploit the geometrical and gate tunability of these nanoresonators to investigate their impact on the photocurrent spectrum and the polaritonic's waveguided modes. This work opens a venue for studying this highly tunable and complex hybrid system, as well as for using it in compact platforms for sensing and photodetection applications.\\
}

Polaritons are coupled excitations of electromagnetic waves with charged particles (plasmons polaritons)\cite{Low2014d, Basov2020} or lattice vibrations (phonon polaritons)\cite{Caldwell2014a, Basov2016a, Giles2018a}. The polaritonic properties become extreme in two-dimensional (2D) materials, including wavelength confinement by factors up to 300,\cite{Iranzo2018b, Tamagnone2018a} ray-like propagating modes,\cite{HerzigSheinfux2024, Dai2015c, Dai2019h} long lifetimes\cite{Taboada-Gutierrez2020, Ni2021a} and capabilities to tune its properties in-situ.\cite{Basov2016a, Sternbach2021} These polaritons have been investigated in near-field studies (e.g. using scanning near-field optical microscopy)\cite{Woessner2014, Woessner2017a, Lundeberg2017a, Lundeberg2017e, Alonso-Gonzalez2017, Dai2015c, Dai2015e, Ni2018, HerzigSheinfux2024, Moore2021}, by electron energy loss spectroscopy (EELS)\cite{Lin2017b, Wachsmuth2014, Li2021c} and far-field with Fourier Transform Infrared spectroscopy (FTIR),\cite{Brar2013c, Brar2014, Low2014d, Rodrigo2015d, Autore2018, Iranzo2018b, Kim2018a, Bylinkin2019, Lee2019c, Epstein2020d, Hu2019a, Lee2019c, Lee2020} which, however, constitute bulky systems that require a typical cooled external detector. In order to achieve a highly compact platform, 2D polaritons have been electrically detected by using a graphene nanodisk array\cite{Guo2018a} or antennas that launch hyperbolic phonon polaritons (HPPs) of hBN in the detector's photoactive area.\cite{Castilla2020} However, the detection is mainly based on increasing the magnitude of the photoinduced signal at a fixed incident wavelength,\cite{Freitag2013g, Bandurin2018e, Guo2018a, Safaei2019} at the expense of spectral information.
\\

Electrical spectroscopy of polaritonic nanoresonators is a novel and unique capability, with prospects for nano-optoelectronic circuits, and molecular sensing applications,\cite{Bareza2020a, Hu2019a} since they can be strongly coupled to molecular vibrations.\cite{Rodrigo2015d, Autore2018, Bylinkin2021} Here, we merge 2D polaritonic resonators with a graphene pn-junction into a single high-quality 2D-material heterostructure to realize the first electrical spectroscopy of deep subwavelength polaritonic nanoresonators. The quality factor of the nanoresonators plays a key role since high values enable their effective photodetection, in contrast to low-quality factor values in low mobility\cite{Iranzo2018b, Lee2019c, Bylinkin2019, Epstein2020d} or patterned graphene\cite{Brar2013c, Brar2014, Low2014d, Rodrigo2015d, Luxmoore2016, Hu2019a}, resulting in a diminished detection efficiency and spectral resolution. Our approach eliminates the need for an external detector for spectroscopy and leads to device miniaturization. Its small photoactive area (comparable to the hot carriers cooling length of $\sim$0.5 to 1 \textmu m\cite{Castilla2020, Tielrooij2018}) is adequate in converting the incoming light into an electrical signal,\cite{Castilla2020} contrary to FTIR which requires large optically active device areas ($\gtrsim$30$\times$30 \textmu m$^2$) to obtain a reasonable signal-to-noise ratio.\cite{Iranzo2018b, Epstein2020d} 
\\

Our methodology has enabled us to investigate for the first time the contribution of the hybridized modes present at the hBN lower reststrahlen bands (RB) in the photocurrent spectrum, which corresponds to a different type of hyperbolicity (type I) with respect to the typically studied (type II) for the upper RB.\cite{Caldwell2014a, Basov2016a} In fact, at the lower RB spectral range, we have observed the highest Q factors and lateral confinement among the whole investigated mid- and long-wave infrared spectra. The active tunability of the polaritonic nanoresonators' spectral photoresponse is explored by gating graphene. We find that doped graphene under certain conditions also acts effectively as a mirror by partially reflecting the polaritons. This modifies the hybridized modes, adding extra degrees of freedom in tuning the device photoresponse.
\\

\begin{figure*} [t]
   \includegraphics[width=\textwidth]
   {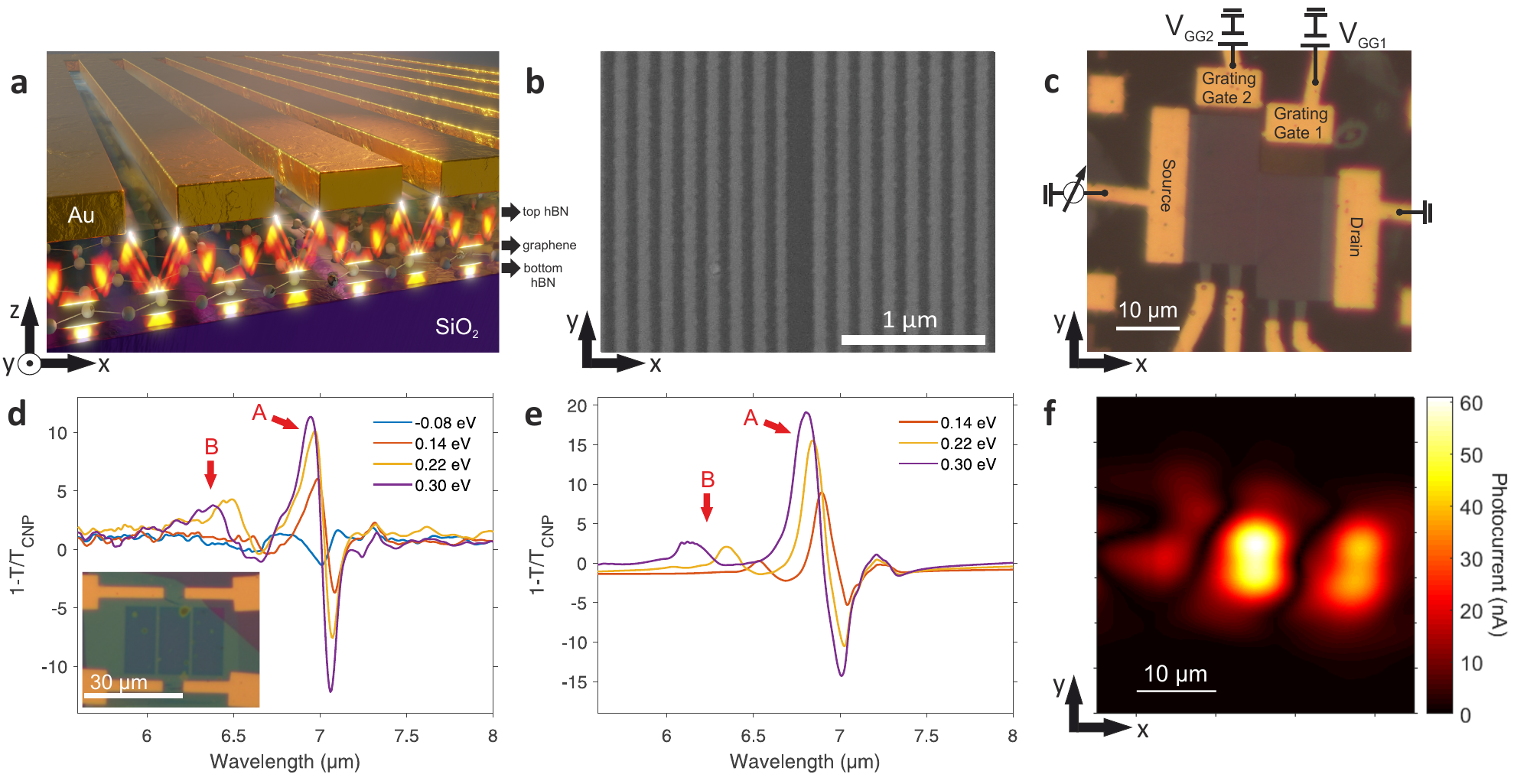}
   \caption{ 
\footnotesize \textbf{Devices' designs, transmission measurements and optical simulations}. \textbf{a)} Schematic representation of the device 1 consisting of metallic nanorods placed on top of a hBN encapsulated graphene. The metal nanostructures provide the necessary momentum for the polaritons to propagate across the 2D stack.
\textbf{b)} SEM picture of the metallic nanorods array with a central gap for devices 2 and 3.
\textbf{c)} Optical image and electrical configuration of the device 3 used for photocurrent measurements.
\textbf{d)} Extinction (1-T$/$T$_{\rm CNP}$) spectrum of the device 1 measured using FTIR. The curves correspond to several Fermi levels as indicated in the legend. The inset shows the optical image of the device 1. The white scale bar corresponds to 30 \textmu m. The three columns above the 2D stack are arrays of 100 nm wide metal nanorods with a 50 nm gap between them. 
\textbf{e)} FDTD simulated extinction spectra of device 1 for several Fermi levels.
\textbf{f)} Scanning photocurrent map (in absolute value) of device 3 at incident wavelength ($\lambda$) of 6.6 \textmu m. The gates are set to GG1 at 0.4 V and GG2 at -0.25 V, thus creating a pn-junction.
}
\label{fig_1}
   \end{figure*}

We investigate four devices with their specifications and fabrication procedure described in Supplementary Note 1 and \textit{Methods}, respectively. Initially, the optical response of these high-quality polaritonic nanoresonators is studied using FTIR, serving as a control experiment. For this purpose, we fabricate devices 1 and 4 that are used exclusively for FTIR measurements owing to their large area requirement (optical active area of $\sim$30$\times$30 \textmu m$^2$, see optical image of device 1 in the inset of Fig. 1d and Supplementary Fig. 1 for device 4) and with a device configuration containing a single backgate to achieve uniform doping in graphene to maximize the optical response (see Supplementary Fig. 2, which indicates that the damping rate decreases with the increase in the Fermi level). However, this gating configuration does not allow the creation of a pn-junction for efficient photodetection.\cite{Castilla2020} Fig. \ref{fig_1}a shows the schematic of devices 1 and 4, which consist of tens of nanometers wide metallic nanorods placed on top of hBN-encapsulated graphene. Upon illumination, scattering at the metallic rod array launches polaritons that propagate across the 2D heterostructure.\cite{Iranzo2018b, Bylinkin2019, Lee2019c, Lee2020, Epstein2020d} The graphene channel is uniformly doped by using a silicon backgate. To obtain a higher yield of fabrication of the metallic nanorods, we pattern them prior to the transfer of the 2D stack in devices 2 and 3, which are used exclusively for photocurrent measurements because they do not require a large optically active area (e.g., device 2 area of $\sim$6$\times$3 \textmu m$^2$, see Supplementary Note 1 for more details). The metal nanostructures of these latter devices have a central gap (shown in Fig. \ref{fig_1}b) to split the array and gate the two graphene regions independently\cite{Castilla2020}, thus creating a pn-junction. However, the grating gates produce a non-uniform electrostatic profile\cite{Delgado-Notario2022, Ryzhii2020, Boubanga-Tombet2020}, which affects the damping rate and optical response as shown in Supplementary Fig. 2. Device 3 is fabricated on an infrared transparent substrate (CaF$_2$) to investigate the hybridized polaritons in more detail since it avoids the presence of phonon polaritons near the hBN RBs spectral regions. In Fig. \ref{fig_1}c, we show an optical image of the device 3 and its electrical circuitry used for photocurrent measurements.
\\

\section*{\textsf{O\lowercase{ptical spectroscopy in the mid-infrared range}}}
Firstly, the optical response of the polaritonic nanoresonators is examined using FTIR to determine the extinction $1-T/T_{\rm CNP}$, where $T$ and $T_{\rm CNP}$ are the transmittances of the device at a certain gate voltage and at charge neutrality point (CNP) respectively.\cite{Iranzo2018b, Epstein2020d, Brar2013c} Fig. \ref{fig_1}d displays the extinction spectra for several Fermi energies. We identify two main peaks that exhibit a graphene plasmonic behavior, which increase their amplitude and blue shift (e.g. $\approx$0.15 \textmu m for peak B) for increasing Fermi level.\cite{Iranzo2018b} Fig. \ref{fig_1}e depicts the simulated extinction using finite-difference time-domain (FDTD) as described in ref. \citenum{Castilla2020} and semi-analytical rigorous coupled-wave analysis (RCWA) described in Supplementary Note 2. We observe excellent qualitative and quantitative agreement with the experimental results that we also support with the dispersion relation of the polaritonic modes present in device 1, as shown in Supplementary Fig. 3, Supplementary Note 2, and by showing the results of device 4 in Supplementary Fig. 5, therefore validating our theoretical model that will be explained in further detail below. The slightly lower experimental values are likely due to peak broadening caused by the inhomogeneity of metal rods' periodicity. This geometrical disorder impacts the scattering time of charge carriers in graphene, thus causing a decrease in peak intensity (see Supplementary Fig. 4). The absorption spatial profiles at the wavelengths of the measured peaks show a hybridized plasmon-phonon polariton for peak A, however, in peak B we do not observe a clear hybridization since the graphene plasmon mode resonates at its plane without interfering with the hBN HPPs (see Supplementary Fig. 6).
\\


\begin{figure*} [t]
   \centering
    \includegraphics[width=\textwidth]
   {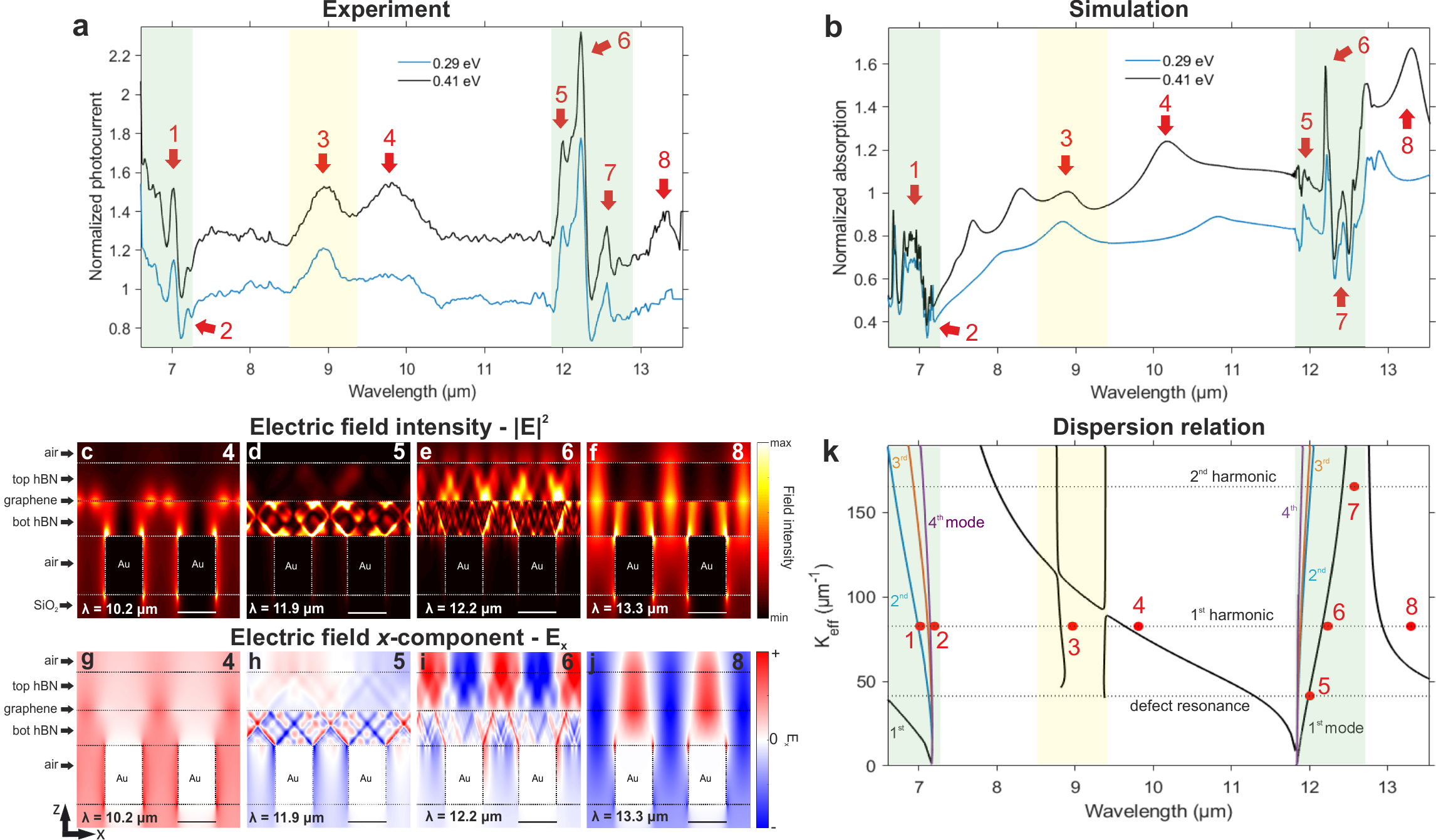}
   \caption{
\footnotesize \textbf{Electrical spectroscopy measurements and simulations} 
\textbf{a)} Normalized photocurrent spectrum of device 2 at several Fermi energies. The photocurrent spectra are normalized to the spectrum at CNP. The polaritonic peaks are labelled by red arrows. The highlighted spectral regions in green correspond to the upper and lower RB of hBN and in yellow to the SiO$_2$ RB. The curves are offset for clarity.
\textbf{b)} Optical (FDTD) simulation of the graphene absorption spectrum for different Fermi energies normalized to the spectrum at CNP. We label the identified peaks in the same manner as the experimental ones in panel \textbf{a}.
\textbf{c-f)} Cross-sectional view of the simulated electric field intensity normalized to the incident one across a region containing two metal nanorods, for wavelengths 10.2, 11.9, 12.2 and 13.3 \textmu m corresponding to peaks 4, 5, 6 and 8 respectively in panel \textbf{a}. The $x-$ (horizontal) and $z-$ (vertical) directions are defined in Fig. \ref{fig_1}a. The white scale bar corresponds to 40 nm. The calculations consider a non-uniform graphene Fermi level with a value of 0.4 eV above the metal (for detailed doping profile see Supplementary Fig. 2).
\textbf{g-j)} Same as panels \textbf{c-f}, but the simulations instead show the cross-sectional view of the $x$-component of the electric field normalized to the incident one. The black scale bar corresponds to 40 nm.
\textbf{k)} Dispersion relation of the polaritonic modes with the respective harmonic diffraction orders (2$\pi$n/D). The three horizontal dashed lines correspond to the defect resonance ($n=1/2$), first ($n=1$) and second ($n=2$) diffraction order resonances launched by the metal rod array, respectively. The marked red dots represent the experimental values, which the numeric labels are defined in Fig. \ref{fig_2}a. The graphene Fermi level is 0.4 eV. At the hBN RBs (green highlighted regions) the black, blue, orange and purple lines correspond to the 1$^{\rm st}$, 2$^{\rm nd}$, 3$^{\rm rd}$ and 4$^{\rm th}$ hybridized polaritonic modes respectively. In yellow is highlighted the SiO$_2$ RB.
 }
\label{fig_2}
   \end{figure*}


\section*{\textsf{E\lowercase{lectrical detection of polaritonic nanoresonators}}}
After determining the optical response and validating the theoretical model, we perform photocurrent spectroscopy measurements for electrical detection of the polaritonic nanoresonators. First, we perform a mid-infrared scanning photocurrent map across the device area. For this, we apply the appropriate voltages to dope the graphene region above the grating gate 1 (GG1) and 2 (GG2) with opposite polarity, creating a pn-junction in the graphene, for which the photocurrent is maximum (see Fig. \ref{fig_1}f and Supplementary Fig. 7). By tuning the gate voltages independently, we observe multiple sign changes of the photocurrent as shown in Supplementary Figs. 8-9, which is consistent with the photothermoelectric (PTE) effect\cite{Castilla2020, Woessner2017a, Lundeberg2017a, Massicotte2021, Viti2021} with a responsivity of $\sim$ 10 \textmu A/W (see the spectral dependence of the responsivity and noise equivalent power in Supplementary Figs. 2 and 10 respectively). Further improvements in responsivity can be made through more optimized designs\cite{Castilla2020, Castilla2019}.
\\

Next, by scanning the wavelength of the source, we spectrally resolve the resonances of the 2D polaritons from the normalized photocurrent spectra of device 2 (see Fig. \ref{fig_2}a). The photocurrent spectra are normalized to the spectrum at CNP to probe the Fermi energy-dependent optoelectronic properties. GG1 is set to a fixed low voltage (doped region) since the Seebeck coefficient is maximum close to CNP\cite{Castilla2020, Lundeberg2017a}, while GG2 is swept towards high negative voltages (doped p-type region), thus creating a doping asymmetry in the channel to maximize the photoresponse. We observe several peaks at the upper ($\approx$ 6-7 \textmu m) and lower ($\approx$ 12-13 \textmu m) RB of hBN and SiO$_2$ RB ($\approx$ 8-9 \textmu m). Some of these peaks at the RBs evolve with Fermi level, which is ascribed to the hybridized plasmon-phonon polaritons. Moreover, two additional broader peaks (labeled as 4 and 8) appear at high Fermi energies outside these RBs. Their evolution and amplitude increase with the Fermi level more pronouncedly than those of the hybridized ones, which is in agreement with previous works.\cite{Brar2013c, Iranzo2018b}
\\

To identify the origin of the resonances in the photocurrent spectrum, we show in Fig. \ref{fig_2}b the simulated normalized absorption using FDTD and RCWA (see Supplementary Fig. 11). The absorption and photocurrent are proportionally related via the electronic temperature gradient\cite{Castilla2020}, as shown in Supplementary Fig. 12. We find an excellent agreement in terms of spectral position and relative amplitude of the peaks. Additional peaks observed in the theoretical curves are due to resonances of higher-order modes.\cite{Iranzo2018b} We investigate the resonances in more detail by analyzing the field distributions above and outside the metal nanorods, as well as between the top and bottom layers of hBN, as shown in Figs. \ref{fig_2}c-f. It can be seen that the field is confined to the bottom hBN above the metal, which corresponds to an acoustic graphene plasmon.\cite{Epstein2020d, Iranzo2018b, Alonso-Gonzalez2017} Conversely, between the metal nanorods, the field extends symmetrically in both hBN layers, as expected for conventional graphene plasmons.\cite{Brar2013c, Woessner2014} Moreover, the electric field distribution (e.g. the $x-$component of the field shown in Fig. \ref{fig_2}g-j) indicates the resonance's harmonic order. For instance, peaks 4 and 8 show one sign change of the field in one period as displayed in Figs. \ref{fig_2}g and \ref{fig_2}j (see also Supplementary Fig. 13), corresponding to the first harmonic. Different harmonic resonances appear at the top and bottom layers of hBN at peak 6 (Fig. \ref{fig_2}i), which implies superposition of the hybridized polaritonic modes. Another interesting case occurs at peak 5, where one period of the field matches two periods of the grating corresponding to a defect resonance (see Fig. \ref{fig_2}h and Supplementary Fig. 14).  
\\


\begin{figure*} [t]
	\centering
	\includegraphics[width=\textwidth]
	{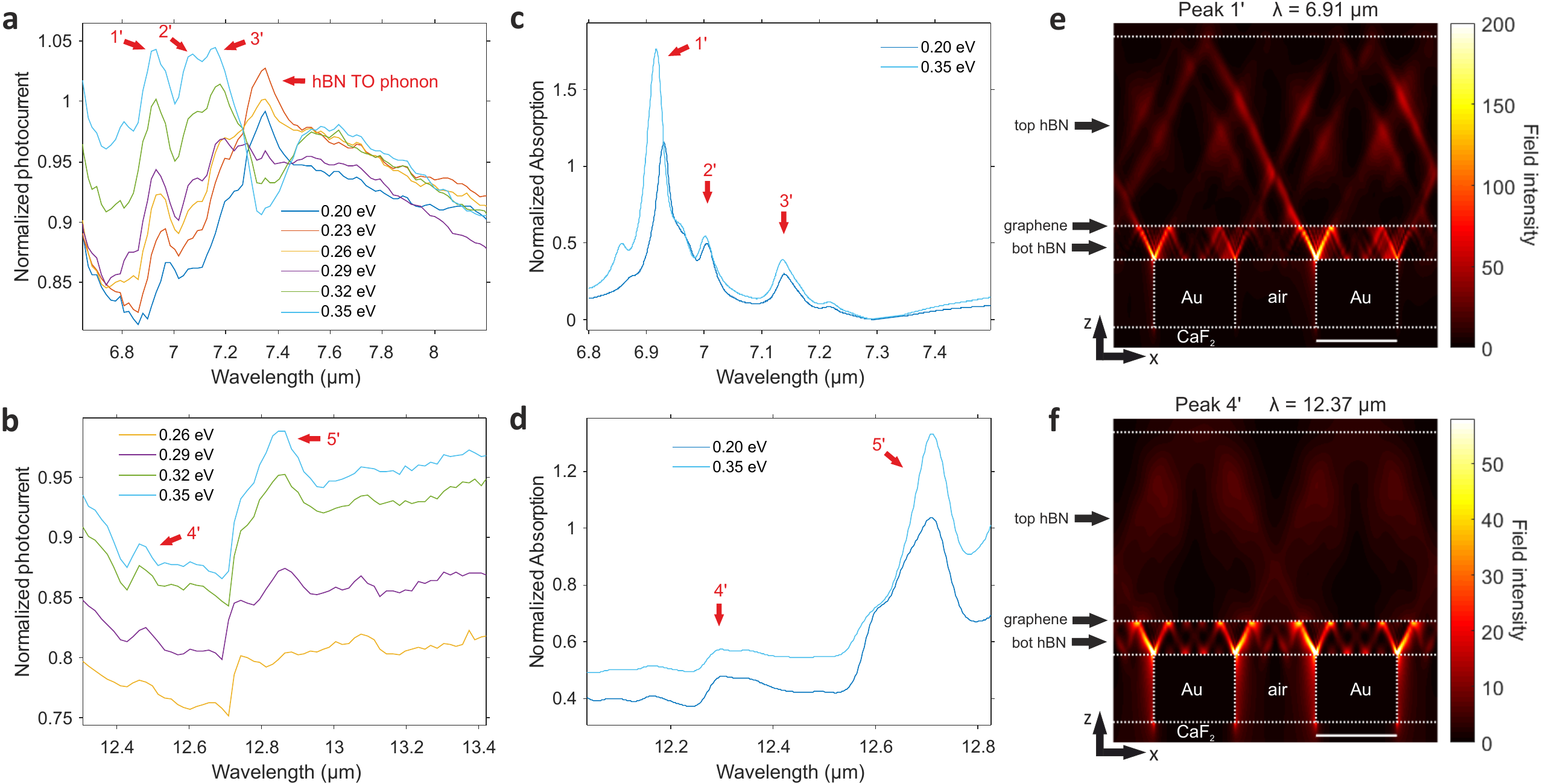}
	\caption{
		\footnotesize \textbf{Normalized photocurrent spectra of device 3 at the hBN RBs.}
		\textbf{a)} Normalized photocurrent spectra at the upper RB of hBN for several gate voltages. The photocurrent spectra are normalized to the spectrum at CNP. The polaritonic peaks are labelled by red arrows. The 0.20 eV curve is slightly shifted (divided by 1.05) for illustration. The Fermi energies are presented in absolute values.
		\textbf{b)} Same as panel \textbf{a} but for the lower RB range.
		\textbf{c)} Optical simulation of graphene absorption at the upper RB spectral region for different Fermi energies normalized to the spectrum at CNP. We label the identified peaks in the same manner as the experimental ones.
		\textbf{d)} Same as panel \textbf{c} but for the lower RB range.
		\textbf{e)} Cross-sectional view of the electric field intensity normalized to the incident one. The $x-$ (horizontal) and $z-$ (vertical) directions are defined in Fig. \ref{fig_1}a. The simulations correspond to a non-uniform graphene Fermi level at 0.35 eV at wavelength 6.91 \textmu m (corresponding to peak 1' in panel \textbf{a}). The white scale bar corresponds to 20 nm.
		\textbf{f)} Same as panel \textbf{e} but for lower RB range at wavelength 12.37 \textmu m (corresponding to peak 4' in Fig. \ref{fig_3}b).
	}
	\label{fig_3}
\end{figure*}


A complementary assessment involves calculating the dispersion relation of the polaritonic modes present in device 2 by using the transfer matrix method (TMM) described in Supplementary Note 2. The metallic rod arrays provide the in-plane effective momentum given by: $k_{\rm eff}^l = k_{\rm x,m}^lw/D+k_{\rm x,g}^lg/D = 2\pi n/D$, where $D$, $w$ and $g$ are the period, width of the nanorod, and the gap between them, respectively. The parameter $l$ represents the order number of the mode and $n\geq1$ is the number of harmonic diffraction orders. The above equation encompasses the combination of both acoustic ($k_{\rm x,m}$, above the metals) and conventional polaritons ($k_{\rm x,g}$, above the gap)\cite{Lee2019c, Lee2020} as described in Supplementary Note 2 and Supplementary Fig. 15. The number of nodes present in the polaritonic field along the vertical direction determines $l$, while the number of nodes of the field along the lateral direction over a period determines $n$ (see also Supplementary Figs. 13 and 16). Noteworthily, $l =$ 0 outside the RBs. We find a generally excellent correspondence with the experimental results as shown in Fig. \ref{fig_2}k. In the lower RB, for simplicity we consider the fundamental mode as the dominant, however, the polaritonic field distribution suggests a superposition of the fundamental with higher-order modes.
\\

\section*{\textsf{G\lowercase{ate tunability of the hybridized polaritonic nanoresonators}}}
The hybridized polaritonic nanoresonators are investigated in more detail with device 3, which contains an infrared transparent substrate (CaF$_2$) to avoid phonon polaritons in the substrate near the hBN RBs spectral range, in particular at the lower RB ($\sim$12 \textmu m).\cite{Kischkat2012b} Following a similar procedure and gates' configuration explained previously, we tune GG1 region to a fixed low n-type doping and sweep GG2 towards high p-type doping. At the upper RB (Fig. \ref{fig_3}a), we identify several peaks of photocurrent whose amplitude increases and their spectral positions blue shift (e.g. 70 nm or 15 cm$^{-1}$ for peak 1') with the increase of Fermi level. In fact, at the highest doping (0.35 eV) the value of normalized photocurrent at peaks 1'-3' exceeds the values of the CNP curve, where the Seebeck coefficient is higher,\cite{Castilla2020} thus showing the absorption enhancement caused by the polaritonic resonances. At the spectral location of the TO phonon of hBN, we observe a pronounced contribution at low Fermi level values which reduces its effect at high doping.\cite{Badioli2014e} In the lower RB (Fig. \ref{fig_3}b), we observe two main peaks that boost their amplitudes at higher Fermi levels, as well as a small blue shift (e.g. 50 nm or 3 cm$^{-1}$ for peak 5'), corroborating the theoretical prediction shown in Supplementary Fig. 17. 
\\

The photocurrent peaks' locations agree well with the peaks in the simulated absorption spectra, shown in Fig. \ref{fig_3}c-d. We point out a small redshift of $\sim$0.15 \textmu m for the spectral position of peaks at the lower RB, as well as a slight broadening of the experimental peaks compared to the theoretical ones. The latter is given by the inhomogeneity of the metallic rods' periodicity as explained previously. Fig. \ref{fig_3}e-f show that the maximum field occurs at the bottom hBN, which is between graphene and metallic nanorods followed by partial transmission of HPPs rays towards the upper hBN. Hence, we have two different types of field distributions in each hBN layer, which are affected by the graphene doping. The dispersion relation of these hybrid polaritonic modes are shown in Supplementary Fig. 18 and demonstrate excellent agreement with experimental results.
\\

\begin{figure*} [t]
	\centering
	\includegraphics[width=\textwidth]
	{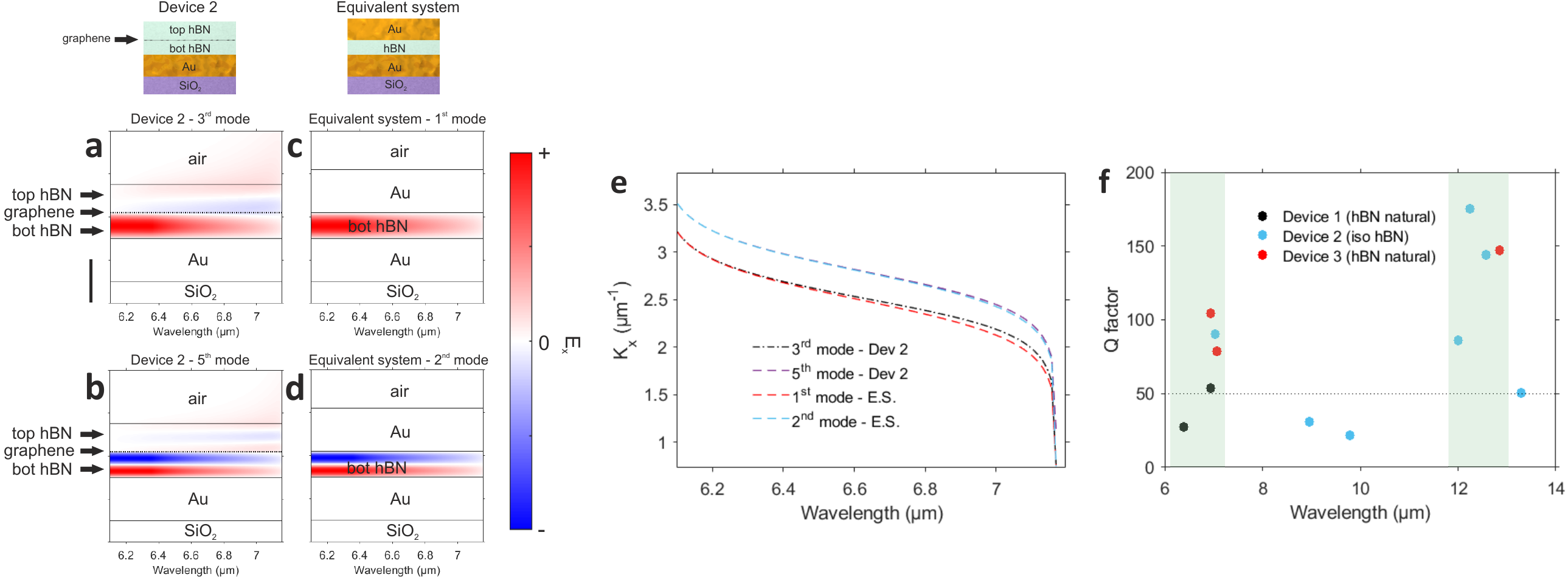}
	\caption{
		\footnotesize \textbf{Spatial field distribution and dispersion of hBN equivalent geometry and Q-factor spectrum of the 2D polaritonic nanoresonators.} 
		Cross-sectional view of the spatial $x-$component of the electric field as a function of the wavelength for \textbf{a)} 3$^{\rm rd}$ and \textbf{b)} 5$^{\rm th}$ mode of the hybridized polariton. The graphene doping is 0.4 eV. The vertical black scale bar corresponds to 10 nm for the panels \textbf{a-d}. \textbf{c)} and \textbf{d)} correspond to the first and second order mode respectively of the bottom hBN (6 nm thick) of the equivalent system (E.S.) geometry consisting on the bottom hBN embedded by two gold layers of 10 nm without the presence of graphene and the top hBN as shown in the illustrations on the top part of the figure.
		\textbf{e)} Dispersion relation of the investigated modes of the two systems.
		\textbf{f)} Q-factor spectrum of the measured 2D polaritonic nanoresonators. The regions highlighted in green correspond to the RBs of hBN.
	}
	\label{fig_4}
\end{figure*}

To analyze the nature of this hBN-confined mode, which is controlled by the graphene Fermi energy, we compare two systems at the upper RB. The first one corresponds to the structure described earlier for device 2 (considering polaritons only above metal) shown in Figs. \ref{fig_4}a-b, which include the spectral and spatial distributions of eigenmodes for the third and fifth modes respectively. We notice that the third mode is characterized by two nodes in the vertical direction, while the fifth one has four nodes. In both cases, one of the nodes is located in the vicinity of graphene. Additionally, the fields are mainly confined in the bottom hBN for both cases. The second system consists of replacing the graphene and top hBN layer with a gold film as shown in Figs. \ref{fig_4}c-d along with their respective spatial distributions of fields from the two first eigenmodes. We observe similarities between the field distributions in the bottom hBN layer and those of the analogous geometry shown in Figs. \ref{fig_4}a-b. This trend is further corroborated by comparing the dispersion curves of both systems, which overlap (see Fig. \ref{fig_4}e). This model demonstrates that doped graphene acts as a mirror, partially reflecting polaritons in the bottom hBN layer.
\\

Two of the key figures-of-merit for this resonant polaritonic system are the Q-factor and the mode volume or wavelength compression, as they are relevant for enhancing the light-matter interactions, for example for sensing applications. However, the quality factors diminish significantly when shrinking the dimensions of the nanoresonators to aim for deep subwavelength confinement.\cite{HerzigSheinfux2024, Lee2020} The presented polaritonic platform overcomes this limitation by showing simultaneously resonances with values up to Q$\sim$200 inside the RBs (see Fig. \ref{fig_4}f), due to the low loss nature of the hybridized polaritons\cite{Dai2015c, Woessner2014} and a record value of 330 is achieved for the optical lateral confinement or effective refractive index\cite{Lee2020} (n$_{\rm eff} \simeq k_{\rm p}/k_{\rm in}$, see Supplementary Fig. 19). In the case of graphene plasmons, we observe lower Q-factors up to 50, most likely limited by Ohmic losses, the non-uniform electrostatic potential and inhomogeneities in the metal nanorods' periodicity that act as additional scattering centers that reduce the resonant peak linewidth, as shown in Supplementary Fig. 2. Additionally, we observe that Q-factor does not show significant variation as a function of Fermi level in agreement with other studies\cite{Brar2013c} (see Supplementary Fig. 20). We point out that the electrical spectroscopy approach enables the probing of very small nanoresonators ($\sim$30 nm), which is highly challenging for conventional techniques such as s-SNOM due to the resolution limitations imposed by the typical tip diameter ($\sim$50 nm)\cite{HerzigSheinfux2024} or FTIR due to the sizeable area requirement (e.g. large arrays of nanoresonators)\cite{Lee2020,Iranzo2018b,Epstein2020d, Lee2019c} as mentioned previously.
\\


\section*{\textsf{D\lowercase{iscussion}}}
The investigated electro-polaritonic platform for performing mid and long-wave infrared photocurrent spectroscopy can be exploited to enhance photodetectors, hyperspectral and sub-diffraction imaging,\cite{Ju2017c, Shen2022c} and electrical detection of molecular vibrations and gases. This highly compact device leads to one step forward for the CMOS integration of electro-polaritonic platforms.\cite{Goossens2017} These devices can also target other frequencies by tuning the size of the metallic nanorods and using different hyperbolic materials such as $\alpha$-MoO$_3$,\cite{Ma2018a, Matveeva2023} V$_2$O$_3$\cite{Taboada-Gutierrez2020} and black phosphorous. In particular, the in-plane anisotropy of MoO$_3$ could potentially change the interference and hybridization with graphene plasmons, thus modifying its effect on photodetection.\cite{Alvarez-Perez2022}
\\


\section*{Methods}

\subsubsection{Device Fabrication}

The fabrication of devices 1 and 4, used for the transmission measurements, consist on the following: we first exfoliate the top and bottom hBN and the graphene onto freshly cleaned Si/SiO$_2$ substrates, stack them following the Van der Waals assembly technique \cite{Wang2013f,Pizzocchero2016} and transfer the hBN/graphene/hBN stack onto a high resistivity Si/SiO$_2$, which is a 50$\%$ transparent and gating capable substrate at the mid-infrared wavelengths.\cite{Iranzo2018b, Epstein2020d} We then use electron beam lithography (EBL) with a PMMA 950 K resist film to pattern source and drain electrodes and expose the device to a plasma of CHF$_3$/O$_2$ gases to partially etch the Van der Waals stack. Subsequently, we deposit side contacts of chromium (5 nm) / gold (60 nm) and lift off in acetone as described in Ref. \citenum{Wang2013f}. Lastly, we pattern the nanorods with a period of 150 nm using EBL and deposit titanium (2 nm) / gold (8 nm) with a subsequently lift off step in acetone.
\\

Devices 2 and 3 were fabricated and designed for photocurrent measurements. On device 3, we first pattern the grating gates by using EBL and deposit titanium (2 nm) / gold (8 nm) followed by a lift off step in acetone. Alternatively, for device 2 we pattern with Ga FIB (gallium focused ion beam) a thin layer of gold deposited on a Si/SiO$_2$ substrate. The dimensions of the metallic nanorods are described in Supplementary Note 1. Following the previously mentioned procedure we transfer the hBN/graphene/hBN stack onto the grating gates. Then, we pattern the source and drain electrodes with a PMMA 950 K resist film using EBL and exposing the patterned regions to a plasma of CHF$_3$/O$_2$ gases to partially etch the Van der Waals stack. Afterwards, we deposit side contacts of chromium (5 nm) / gold (80 nm) and lift off in acetone as described in Ref. \citenum{Wang2013f}. Then, we define the hBN-encapsulated graphene channel by patterning a PMMA mask with EBL and etch it using a CHF$_3$/O$_2$ plasma. By performing electrical measurements using 2-terminal configuration as a function of the gate voltages (varying GG1 and GG2 both at the same potential), we obtain 3,000-15,000 cm$^2$V$^{-1}$s$^{-1}$ as a lower bound of the estimated mobility (see Supplementary Fig. 2).
\\

\subsubsection{Measurements}
For the transmission measurements of devices 1 and 4 we use a commercial FTIR (fourier transform infrared) spectrometer (Bruker Tensor FTIR with a Bruker Hyperion 2000 microscope) and nitrogen cooled mercury-cadmium-telluride (MCT) detector, which its spectral range goes from 6500 to 650 cm$^{-1}$ ($\lambda =$ 1.54 to 15.4 \textmu m) under normal incidence in air with p-polarized light (i.e. with incident polarization perpendicularly oriented respect to the main axis of the metallic gratings).\cite{Iranzo2018b} We use a spectral resolution of 16 nm (4 cm$^{-1}$). We normalize the transmission spectrum with a reference signal at the graphene area with the backgate at the charge neutrality point (CNP) value ($\sim$0 V).	
\\

For the photocurrent spectroscopy measurements, we use a quantum cascade laser (QCL) mid and long-wave infrared laser (MIRcat from Daylight Solutions) that its tunable wavelength ranges from 6.6 to 13.6 \textmu m with a spectral resolution of <1 cm$^{-1}$ and it's linearly polarized. We modulate the light via an optical chopper at 373 Hz and we measure the photocurrent using a lock-in amplifier (Stanford Research). We scan the device position with motorized xyz-stage. We focus the infrared light using a reflective objective with a NA of 0.5. To calibrate the incident power we use a thermopile detector from Thorlabs placed at the sample location.
\\

\section*{Acknowledgments}
\small
The authors thank Hanan Herzig Sheinfux, Krystian Nowakowski and Iacopo Torre for fruitful discussions. F.H.L.K. acknowledges financial support from the Spanish Ministry of Economy and Competitiveness, through the “Severo Ochoa” Programme for Centres of Excellence in R$\&$D (SEV-2015-0522), support by Fundacio Cellex Barcelona, Generalitat de Catalunya through the CERCA program,  and  the Agency for Management of University and Research Grants (AGAUR) 2017 SGR 1656.  Furthermore, the research leading to these results has received funding from the European Union Seventh Framework Programme under grant agreement no.785219 and no. 881603 Graphene Flagship for Core2 and Core3. S.C. acknowledges financial support from the Barcelona Institute of Science and Technology (BIST), the Secretaria d’Universitats i Recerca del Departament d’Empresa i Coneixement de la Generalitat de Catalunya and the European Social Fund (L'FSE inverteix en el teu futur) – FEDER. N.M.R.P. acknowledges support from the Independent Research Fund Denmark (grant no. 2032-00045B) and the Danish National Research Foundation (Project No. DNRF165). K.W. and T.T. acknowledge support from the JSPS KAKENHI (Grant Numbers 21H05233 and 23H02052) and World Premier International Research Center Initiative (WPI), MEXT, Japan for the growth of h-BN crystals. Funding for hBN crystal growth by E.J. and J.H.E. was provided by the Office of Naval Research, Award no. N00014-22-1-2582.
\section*{Author contributions}
\small
S.C, D.A.I. and F.H.L.K. conceived the project. S.C. and H.A. fabricated the devices and performed the experiments. D.A.I. and A.G. assisted in the experiments. M.C. and R.K.K. supported the device fabrication. I.V., Y.B., N.M.R.P. and E.L. performed the simulations and developed the theoretical model. S.C. and M.I.V. assisted in the modelling. S.C., I.V., Y.B., E.L and F.H.L.K. wrote the manuscript. K.W. and T.T. synthesized the hBN crystals. E.J. and J.H.E. synthesized the isotopically enriched hBN crystals. N.M.R.P., E.L. and F.H.L.K. supervised the work and discussed the results. All authors contributed to the scientific discussion and manuscript revisions. S.C., H.A., I.V. and Y.B. contributed equally to the work.

\clearpage

\clearpage

\newpage


\clearpage

\newpage


\renewcommand\refname{Supplementary references}
\def\bibsection{\section*{\refname}} 

\renewcommand{\figurename}{Supplementary~Figure}
\renewcommand{\tablename}{Supplementary~Table}
\setcounter{equation}{0}
\setcounter{figure}{0}
\setcounter{table}{0}
\setcounter{page}{1}

\section*{Supplementary Note 1: Description of the devices characteristics}

We produced four devices in total. Owing to the different requirements, we use two device configurations depending on the type of measurements, by mainly varying the grating location. We fabricated two devices that are exclusively for FTIR and two others for photocurrent measurements. For FTIR, we required a large optically active area ($\sim$30x30 \textmu m$^2$) of the hBN-encapsulated graphene combined with metallic nanorods on top of the 2D stack for devices 1 and 4, as shown in Fig. 1a of the main text, Supplementary Fig. 1a-c, and the inset of Fig. 1d in the main text. This configuration provides an ideal situation for launching efficiently the polaritons \cite{Iranzo2018b, Epstein2020d} and achieving uniform gating across the graphene channel by using a Si backgate, which enhances the optical response of the polaritonic nanoresonators by obtaining the same low damping rate across the graphene channel with a uniform gating profile (see Supplementary Fig. 2c-d, which shows that the damping rate decreases with the increase of the Fermi level). However, owing to this configuration of a single backgate, we are not able to produce a proper graphene pn-junction, and thus, cannot perform efficient photodetection measurements.
\\

One of the two main drawbacks of this platform is the highly challenging fabrication procedure to obtain a relatively large clean interface of the van der Waals heterostructures and homogeneous metallic nanorods across this large area, with the purpose of obtaining a decent signal-to-noise-ratio (SNR) in transmission measurements. In fact, even though achieving the minimum optically active area required, the SNR could still be low, as shown for device 4 results, in Supplementary Fig. 5. The second challenge is the metallic nanogratings arduous lift-off step, as shown in Supplementary Figure 1a, which sometimes requires sonication that harms the graphene and 2D stack quality, hence obtaining a quite low  fabrication yield.
\\

\begin{figure*}[h!]
	\includegraphics [width=\textwidth,scale=0.40]
	{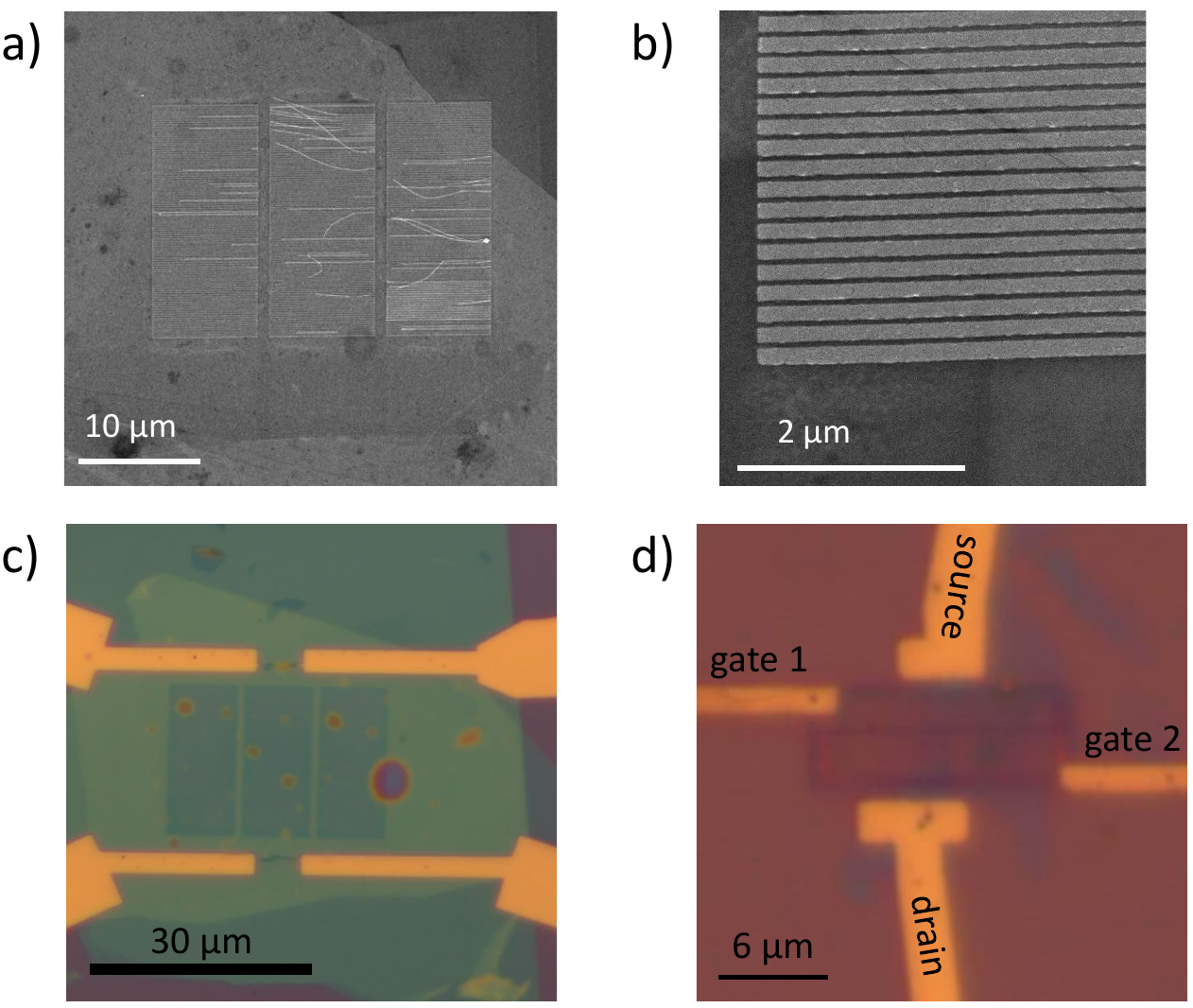}
	\caption{ 
		\footnotesize \textbf{a)} SEM picture of the gratings on top of the 2D stack of device 1. \textbf{b)} Zoomed SEM image of the gratings in \textbf{a}. \textbf{c)} Optical image of device 4. \textbf{d)} Optical image of device 2. 
	}
	\label{lwir_sem_pics}
\end{figure*}


In order to obtain higher yield of fabrication for these devices, we change the device configuration by placing the gratings below the 2D stack (for devices 2 and 3, see Figs. 1c and Supplementary Figure 1d). In fact, since the first fabrication steps are to pattern and evaporate the metallic gratings, we can sonicate the substrate to avoid any lift-off issues, thus achieving a high yield of fabrication of these nanostructures. Also this alternative device configuration allows us to pattern the metallic gratings of device 3 using Ga FIB (gallium focused ion beam), thus obtaining sharper metallic edges of the gratings and higher resolution of the features, as shown in Fig. 1b in the main text. In contrast to the devices 1 and 4, in devices 2 and 3 we use the gratings with a two-fold purpose: 1) to provide enough momentum to launch the 2D polaritons and 2) to dope graphene by using them as a bottom split gate to create a graphene pn-junction, as shown in Fig. 2. For the photocurrent measurements, the device area can be smaller (e.g. device 2 has an area of 6x2 \textmu m$^2$, see Supplementary Figure 1d) since we could get reasonable SNR in these measurements owing to the small photoactive area to produce a signal in the graphene pn-junction\cite{Castilla2020}. It is worth mentioning that although the grating gates create a graphene pn-junction for efficient photocurrent measurements, these nanostructures produce a non-uniform doping profile, as shown in the previously mentioned figures (Supplementary Fig, 2c-d). Therefore, we conclude that the ideal platform that accomplishes all the requirements to perform simultaneously both transmission and photocurrent measurements is highly challenging to fabricate and is out of the scope of this manuscript.
\\

The 4 devices are listed in Supplementary Table \ref*{table_fab_lwir} with their respective characteristics. We mainly vary the grating period ($L$ is the period of the metallic rods that consists on the sum of the metallic width ($w$) and the gap between these rods ($g$)), substrate, gratings fabrication procedure (fabricated using EBL or FIB), hBN type (natural or enriched isotope) and the graphene channel area of each device.
\\


\newcolumntype{s}{>{\columncolor[HTML]{AAACED}} p{3cm}}
\setlength{\arrayrulewidth}{0.3mm}

\begin{savenotes}
	\begin{table}[ht]
		\begin{threeparttable}
			\centering 		
			\small
			\begin{tabular}{|c|c|c|c|c|c|c|c|}			
				\hline                       
				\textbf{Sample} & \textbf{Grating} & \textbf{Grating} & \textbf{Grating period} & \textbf{Substrate} & \textbf{hBN type} & \textbf{Thickness hBN} & \textbf{Channel area}\\ [0.5ex]
				& \textbf{location} & \textbf{fabrication} & \textbf{($L = w + g$)} &  &  & \textbf{top / bottom} & \textbf{(length $\times$ width)}\\ [0.5ex]
				\hline			                  	
				Device 1 & above 2D stack  & EBL & 150 = 100 + 50 nm & SiO$_2$ & natural & 13 / 4.5 nm & 30 $\times$ 22 \textmu m$^2$\\  
				\hline
				Device 2 & below 2D stack & FIB & 75.5 = 38.5 + 37 nm & SiO$_2$ & iso-B10 & 6.5 / 6 nm & 6 $\times$ 3 \textmu m$^2$\\
				\hline
				Device 3 & below 2D stack & EBL & 100 = 50 + 50 nm & CaF$_2$ & natural & 28 / 5 nm & 24 $\times$ 23 \textmu m$^2$\\
				\hline
				Device 4 & above 2D stack & EBL & 100 = 50 + 50 nm & SiO$_2$ & natural & 12.5 / 8 nm & 30 $\times$ 25 \textmu m$^2$\\
				\hline
			\end{tabular}
			\caption{Characteristics of the fabricated devices.}				
			\label{table_fab_lwir}
		\end{threeparttable}	
	\end{table}
\end{savenotes}



\begin{figure*}[h!]
	\includegraphics [width=\textwidth,scale=0.65]
	{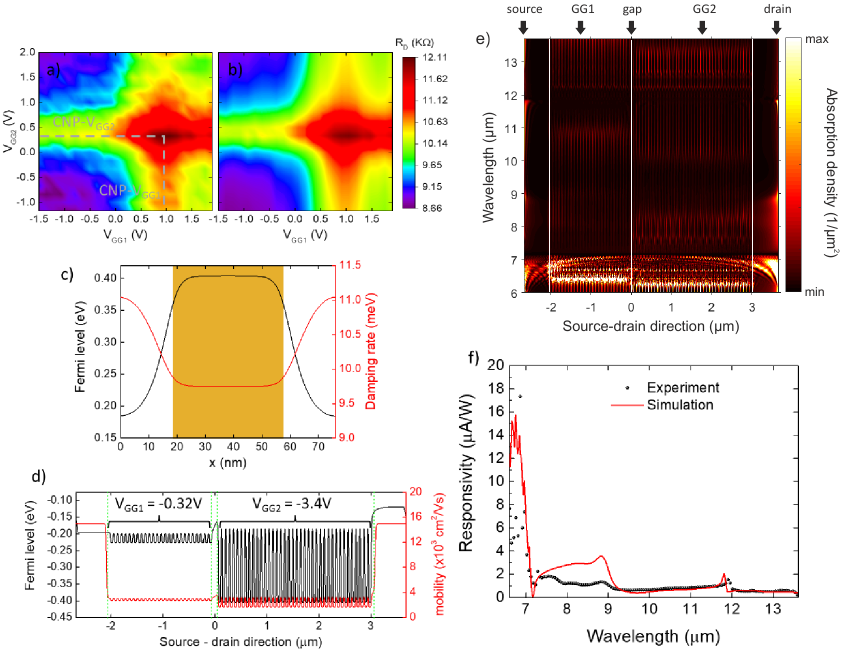}
	\caption{\footnotesize \textbf{a)} Two-terminal experimental and \textbf{b)} fitted resistance as a function of the two grating gates voltages (GG1 and GG2) for Device 2. The resistance was fitted using the model described in ref. \citenum{Castilla2020}, yielding an average field effect mobility $\textless\mu\textgreater \sim$ 3,000 cm$^2/$Vs contact resistance $R_C = 7.2$ k$\Omega$, residual doping of 11.3$\times$10$^{11}$ cm$^{-2}$ at the graphene above GG1 and 6.8$\times$10$^{11}$ cm$^{-2}$ at the graphene above GG2 and V$_{\rm CNP}$ of 0.92 V and 0.34 V for GG1 and GG2, respectively. \textbf{c)} Non-uniform electrostatic potential profile (black curve) and damping rate (red curve) in a grating period for |V$_{\rm GG2}| = 3.4$ V. The yellow shaded region indicates the position of the Au grating gate. The damping rate of the graphene was modeled as $\dfrac{\hbar}{\tau_{\rm MSC}} = \dfrac{\hbar}{\gamma|E_{\rm F}|}$ for the ungated regions near the contacts, while at the gated regions was modified to $\frac{\hbar}{\tau_{\rm MSC}} = \hbar(\frac{1}{\gamma|E_{\rm F}|}+\frac{v_{\rm F}}{\textless A \textgreater})$, with $\textless A \textgreater$ the average length where inhomogeneity in the metal and gap width of the rods appears, to account for an extra scattering mechanism that arises from the device’s geometry disorder. $\gamma$ was estimated at 1500 fs/eV and $\textless A \textgreater = D$ with $D$ the grating period. \textbf{d)} E$_{\rm F}$ (black curve) and $\mu$ (red curve) profile across the graphene channel for V$_{\rm GG1}$ = -0.32 V and V$_{\rm GG2}$ = - 3.4 V. Above GG1 and GG2 $\textless\mu\textgreater \sim$ 2,500 cm$^2/$Vs, while in the ungated region in the vicinity of the contacts $\mu =$ 15,000 cm$^2/$Vs. The mobility was modeled as $\mu = \frac{e\tau_{\rm MSC}v^2_f}{|E_{\rm F}|}$. \textbf{e)} Graphene absorption density $(1/\mu m^2)$ of Device 2 as a function of wavelength across the source drain direction (see Fig. 1a-c in the main text for axis definition, where x = 0 is the center of the gap between the grating gates as indicated with the white vertical line). In panel \textbf{e}, we show the spatially resolved absorption spectrum in graphene across the channel in the $x-$direction at the graphene doping of panel \textbf{d}. We notice different absorption peaks in the GG2 region compared to the GG1 one, where spectrally appear additional absorption peaks outside the RBs range. Also, at the RBs, we observe change of the spectral shape of the absorption peak due to the graphene doping. The end of each gate is indicated with the outer white lines. The absorption contribution at the spatial edges of the device occurs at the ungated graphene region in the vicinity of the contacts (at the left and right corners of the plot) and spectrally located mainly at the RBs of hBN and SiO$_2$. \textbf{f)} Experimental (black dots) and theoretical (red solid line) spectral external responsivity of the device 2 for GG1 at -0.32 V and GG2 at -3.4 V. The typical incident power ranges from $\sim$1-15 mW and an irradiance of 33.6 mW/\textmu m$^2$. The devices 2 and 3 achieve a similar responsivity in the order of tens of \textmu A/W. The low responsivity of these devices relies on the non-optimized design to exploit efficiently the PTE effect as shown in other studies\cite{Castilla2020, Castilla2019}. We point out that the non-uniform graphene doping profile adds more complexity to the photoresponse modeling (see panels \textbf{c-e}) since the absorption of the periodic structure of the two gated regions contribute to the photoresponse besides the absorption around the graphene pn-junction interface\cite{Castilla2020}, in particular for small devices (device 2).
	}
	\label{non-uniform}
\end{figure*}



\begin{figure*}[h!]
	\includegraphics [scale=0.7]
	{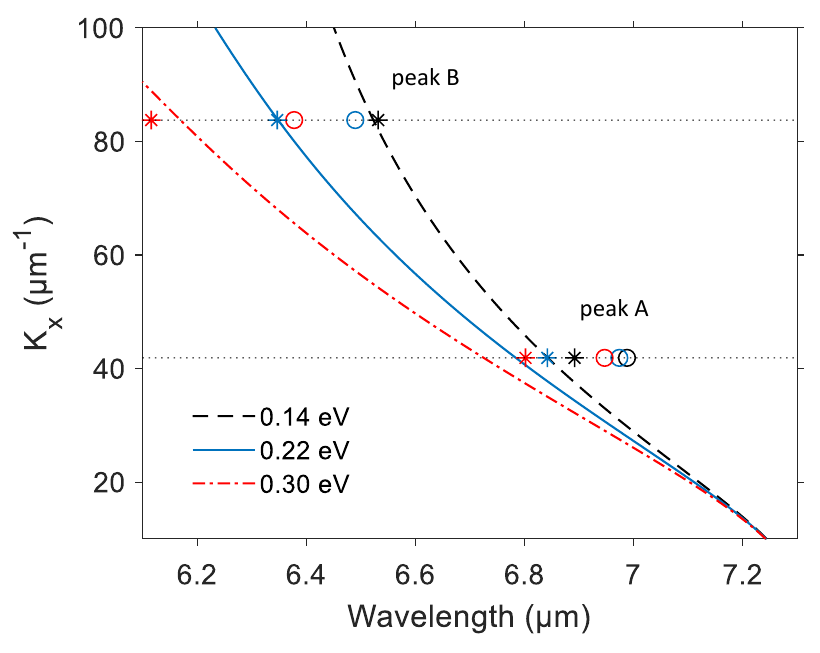}
	\caption{\footnotesize Dispersion relation for device 1 of the hybridized plasmon phonon polariton modes at the upper RB of hBN for different Fermi levels (represented as lines in the legend). The two horizontal dashed lines correspond to the first and second diffraction order resonances launched by the metal rod array. The open dots and asterisks represent the experimental and theoretical values, which correspond to peaks A and B, as defined in Fig. 1d-e in the main text. The slight red shift of the experimental points with respect to the theoretical data is ascribed to the intrinsic doping of graphene.
	}
	\label{relation_dev1}
\end{figure*}


\begin{figure*}[h!]
	\includegraphics [width=\textwidth,scale=0.65]
	{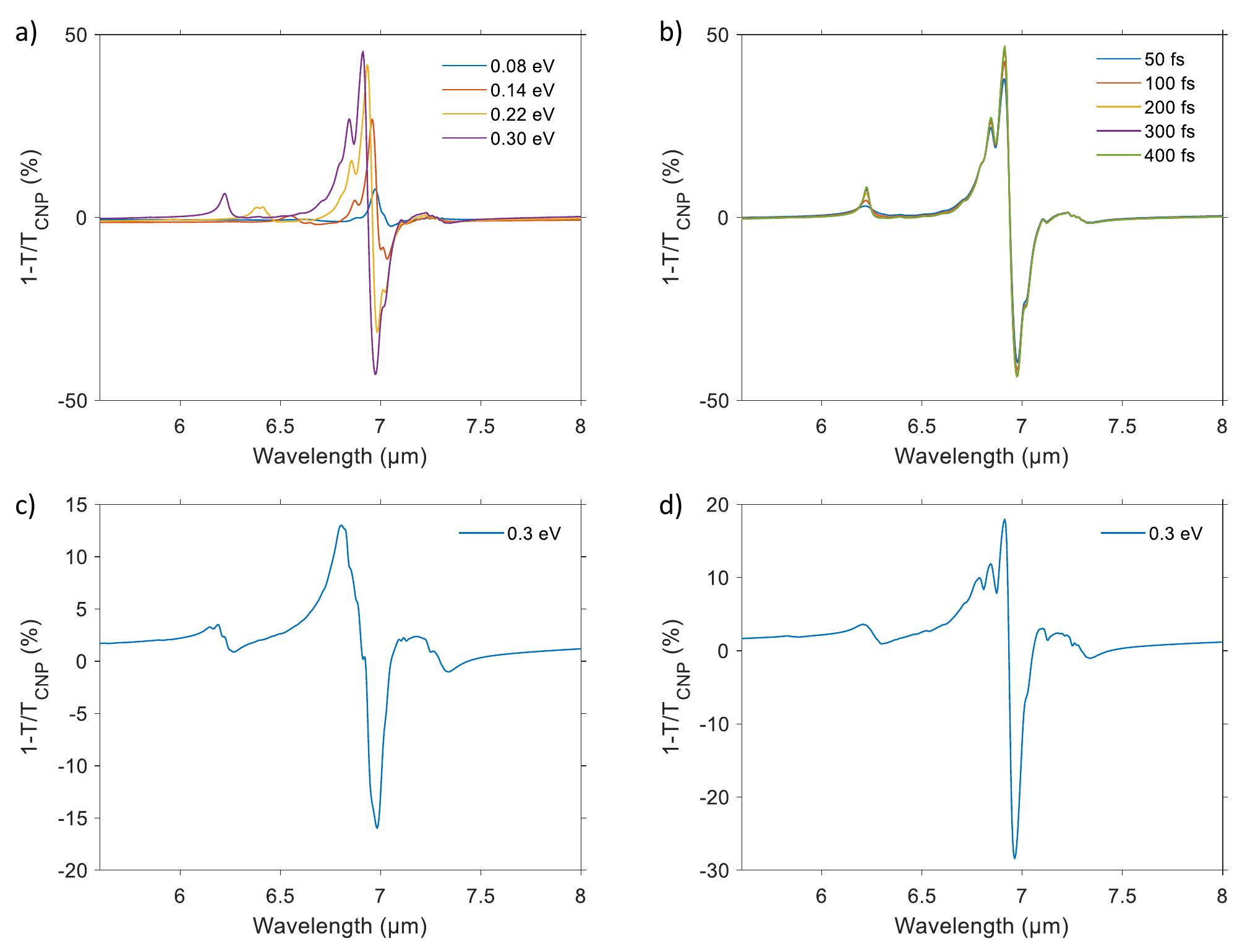}
	\caption{ 
		\footnotesize Extinction simulations of device 1. \textbf{a)} Extinction spectra for different Fermi levels for a fixed scattering time of 200 fs. \textbf{b)} Extinction spectra for several scattering times for a fixed Fermi level of 0.3 eV. \textbf{c)} Extinction spectrum for a fixed Fermi level of 0.3 eV and scattering time of 200 fs. Here the extinction spectrum is calculated for a size distribution of the gap between the metals ranging from 30 to 70 nm. \textbf{d)} Same as \textbf{c} but the extinction spectrum is calculated for a size distribution of the metal width ranging from 80 to 120 nm. All the simulations consider a graphene mobility of 10,000 cm$^2$/Vs. The extinction spectra in \textbf{c} and \textbf{d} are calculated as follows: $E_d = \sum_{i=a_0-N/2}^{i= a_0+N/2} E_i f(i; a_0,\sigma^2)$, where  $E_d$ the extinction spectra in the disorder case, $a_0$ the central value of the metal width (gap width) distribution in \textbf{c} (\textbf{d}), $N$ the extent of the distribution, $E_i$ the extinction spectra for each $i$ metal width (gap width) and $f$ a probability function centered at $a_0$ with standard deviation $\sigma$ = 6 nm.
	}
	\label{abs_broadening}
\end{figure*}


\begin{figure*}[h!]
	\includegraphics [width=\textwidth,scale=0.65]
	{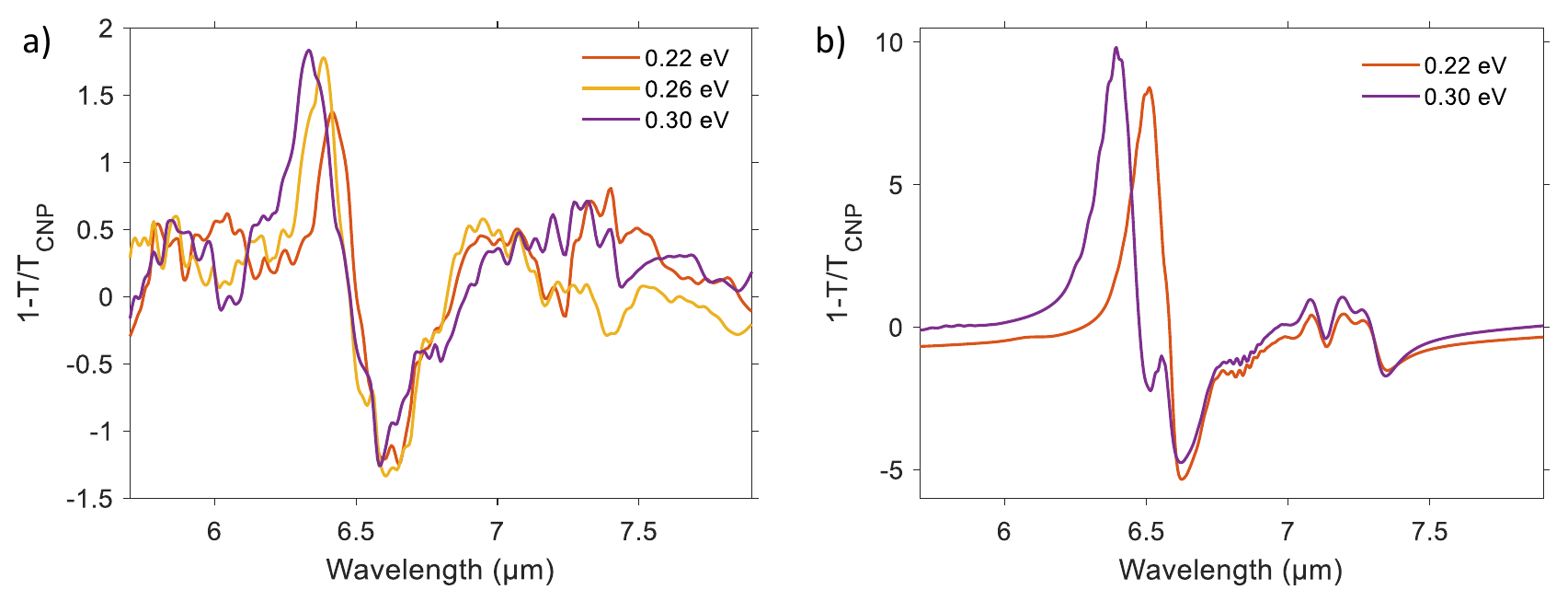}
	\caption{ 
		\footnotesize Extinction measurements and simulations of device 4. \textbf{a)} Extinction (1-T$/$T$_{\rm CNP}$) spectrum of device 4 measured using FTIR. The curves correspond to several Fermi levels, as indicated in the legend. Despite the low extinction values, we are able to probe the polaritonic resonance. \textbf{b)} Simulated extinction spectra of device 4 for several Fermi levels. We use a graphene mobility of 10,000 cm$^2$/Vs. As explained previously, we use a size distribution of the gap between the metals with a standard deviation $\sigma$ = 10 nm.
	}
	\label{device_4_ftir}
\end{figure*}


\begin{figure*}[h!]
	\includegraphics [width=\textwidth,scale=0.65]
	{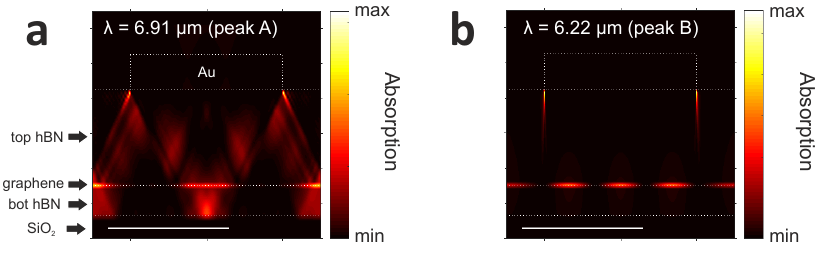}
	\caption{ 
		\footnotesize Absorption profiles of transmission measurements of device 1. 
		\textbf{a)} Cross section view of the simulated absorption of the 2D stack at $\lambda =$ 6.91 \textmu m (1447 cm$^{-1}$) for a graphene Fermi level of 0.3 eV. The white scale bar corresponds to 50 nm.
		\textbf{b)} Same as \textbf{f} but at $\lambda =$ 6.22 \textmu m (1607 cm$^{-1}$). \textbf{a} and \textbf{b} $y-$axis is not to scale for illustration.
		For peak A shown in panel a, we observe that the hBN hyperbolic phonon polaritons (HPPs) are launched at the edges of the metallic gratings and propagate as ray-like waves through the hBN slab. Simultaneously, in the graphene layer plane, a plasmonic wave is observed that also constructively interfere with the HPPs. This behavior corresponds to the hybridize plasmon-phonon polaritons as shown in near-field experiments via s-SNOM \cite{Dai2015c, Woessner2014}. This is in agreement with the relatively small wavenumber shift ($\approx$ 20 cm$^{-1}$) with the Fermi level as shown in Fig. 1d in the main text and is ascribed to the phonon-like nature of this polariton.\cite{Dai2015c, Woessner2014} On the other hand, for peak B (see panel b), we don't observe a clear hybridization since the graphene plasmon mode resonates at its plane without hBN HPPs interferences. This might be due to the fact that this peak is spectrally located at the edge of the upper RB, where the HPPs angles become more steep (close to $\sim$90º).\cite{Caldwell2015a} We point out that peak A shows a narrower linewidth (see Fig. 1d-e in the main text) and higher extinction value ascribed to the low-loss nature of this hybridized polariton \cite{Dai2015c, Woessner2014} in comparison with the peak B. 
	}
	\label{absorption_prof}
\end{figure*}


\begin{figure*}[h!]
	\includegraphics [width=\textwidth,scale=0.65]
	{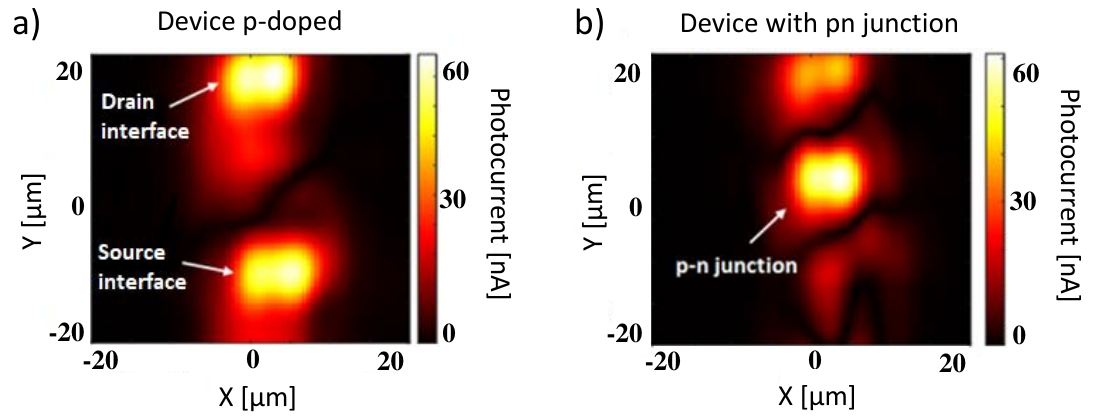}
	\caption{ 
		\footnotesize Scanning photocurrent map as function of the $x$ and $y$ motorized stage position at incident wavelength of 6.6 \textmu m for \textbf{a)} uniformly p-doped graphene channel with both gates (GG1 and GG2) set at -0.2 V and for \textbf{b)} pn-junction graphene channel configuration with GG1 at 0.4 V and GG2 at -0.25 V. The measurements correspond device 3. The photocurrent signal corresponds to the absolute value of the signal without considering the sign change of the photocurrent in vicinity to the metal electrodes.  
	}
	\label{lwir_sem_pics}
\end{figure*}


\begin{figure*}[h!]
	\includegraphics [width=\textwidth,scale=0.65]
	{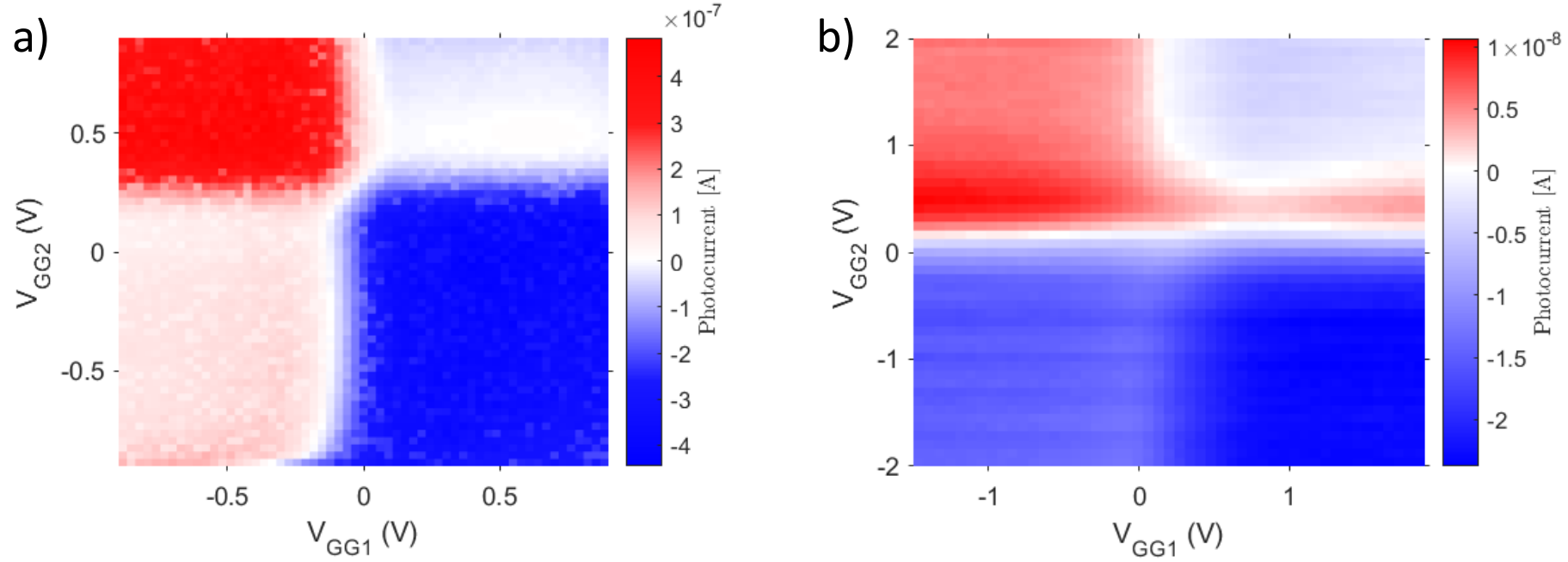}
	\caption{ 
		\footnotesize Photocurrent map as a function of the grating gate voltages (GG1 and GG2) for \textbf{a)} device 3 at incident wavelength of 7 \textmu m and \textbf{b)} device 2 at incident wavelength of 7.75 \textmu m.
	}
	\label{lwir_sem_pics}
\end{figure*}


\begin{figure*}[h!]
	\includegraphics [width=\textwidth,scale=0.65]
	{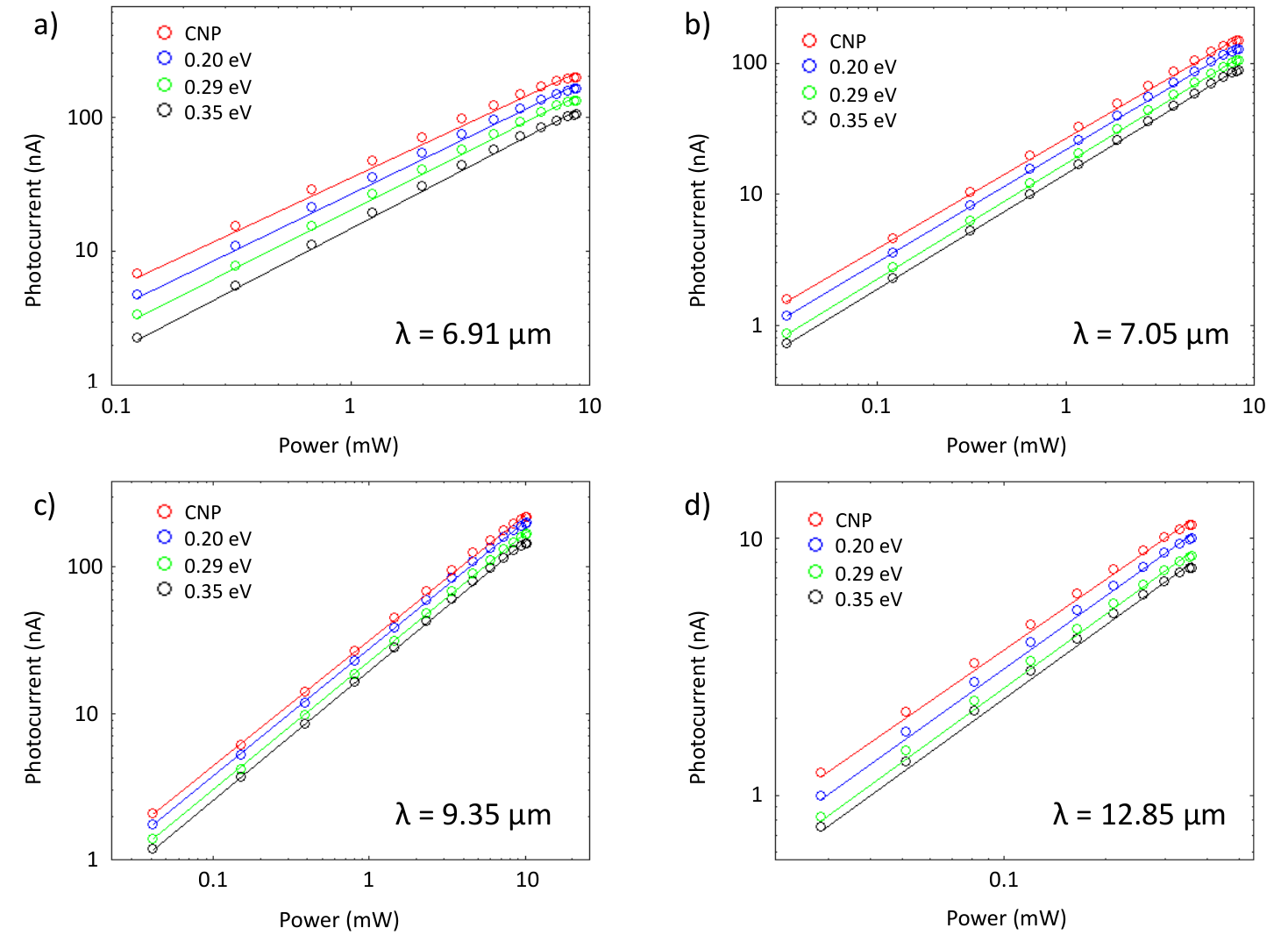}
	\caption{ 
		\footnotesize Power dependence of the photoresponse of device 3. Photocurrent as a function of the incident power for several wavelengths: \textbf{a)} at 6.91 \textmu m, \textbf{b)} at 7.05 \textmu m, \textbf{c)} at 9.35 \textmu m and \textbf{d)} at 12.85 \textmu m. For all the cases we plot for different gate voltages of GG2 that 0 V corresponds to the CNP and -2.4 V to a graphene Fermi level of 0.35 eV. GG1 remains fixed at 0.5 V. The opened circles represent to the experimental points and the lines to the fit that show a linear dependence. This behavior corresponds to the weak heating regime as shown in previous studies.\cite{Tielrooij2018, Castilla2020}
	}
	\label{lwir_8}
\end{figure*}



\begin{figure*}[h!]
	\includegraphics [scale=0.65]
	{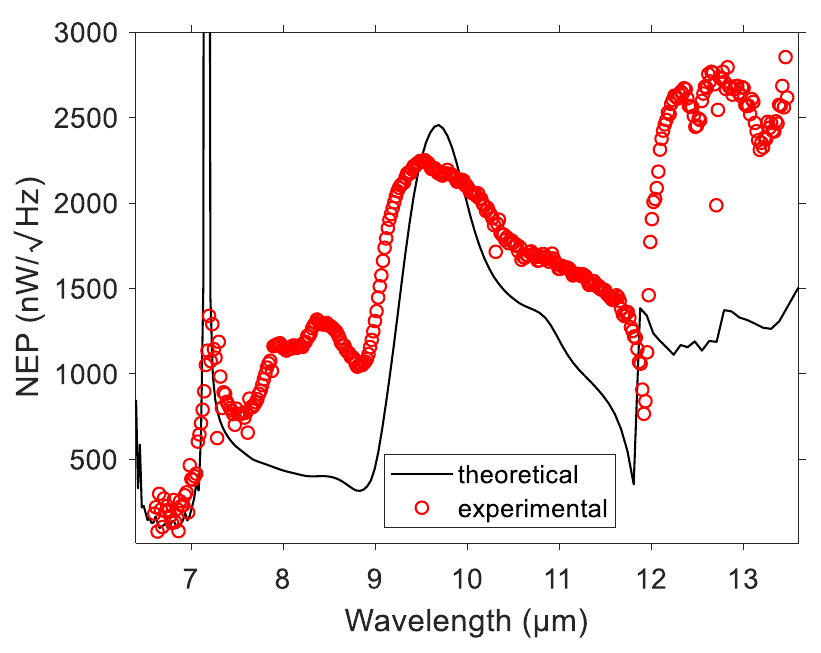}
	\caption{\footnotesize Theoretical (black solid line) and experimental (red open circles) noise equivalent power (NEP) spectrum of device 2. Due to the zero-bias operation, we consider the Johnson noise as the main contribution of the noise in the graphene channel.\cite{Castilla2020, Viti2021} The resistance value of 8.5 kOhm corresponds to the gates' configuration of GG1 at -0.32 V and GG2 at -3.4 V. The minimum NEP corresponds to 77 and 94 nW/$\sqrt{Hz}$ for the experimental and theoretical respectively at $\sim$6.64 \textmu m.
	}
	\label{NEP_sensitivity}
\end{figure*}


\begin{figure*}[h!]
	\includegraphics [width=\textwidth,scale=0.65]
	{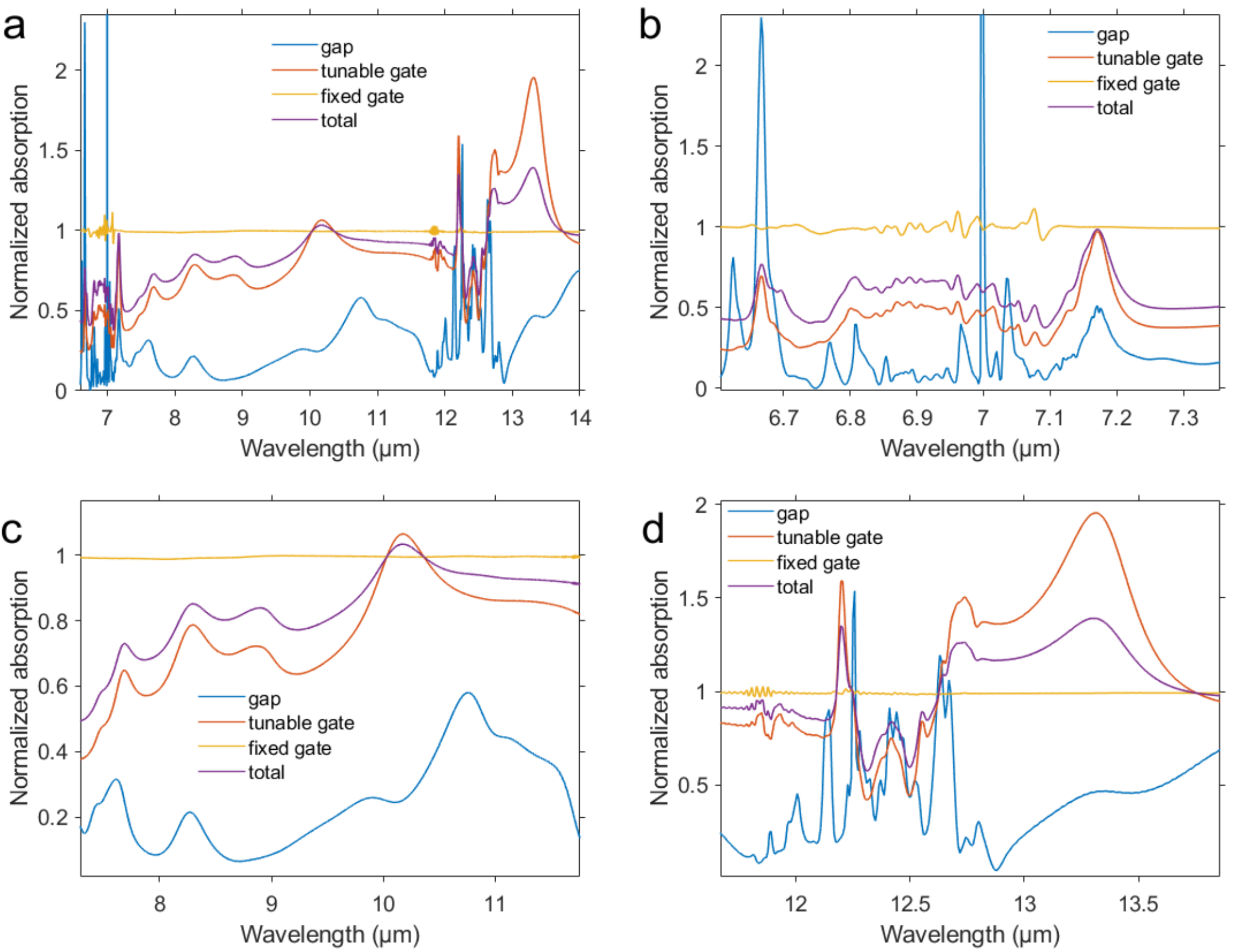}
	\caption{ 
		\footnotesize Simulated graphene absorption spectrum at 0.41 eV normalized to the spectrum at CNP of device 2 for \textbf{a)} the experimentally measured wavelength range, \textbf{b)} upper RB range, \textbf{c)} between RBs and \textbf{d)} for lower RB range. We plot the different contributions of the graphene absorption in the each region of the graphene channel such as the absorption spectrum of the region above the grating gate 2 (GG2) that corresponds to the tunable gate normalized the spectrum at the CNP of that region, the absorption at the region above the gap of the gates normalized to the spectrum of the CNP of the gap region, and the total absorption, which the latter consists on the integrated absorption across the whole graphene channel including all the regions. The region above the grating gate 1 (GG1) that has a fixed Fermi energy (0.21 eV) is normalized to the same spectrum in order to verify the effect of the tunable gate, which we observe that the absorption above the GG1 region is not affected by the change of Fermi energy of GG2 region.
	}
	\label{lwir_sem_pics}
\end{figure*}



\begin{figure*}[h!]
	\includegraphics [scale=0.8]
	{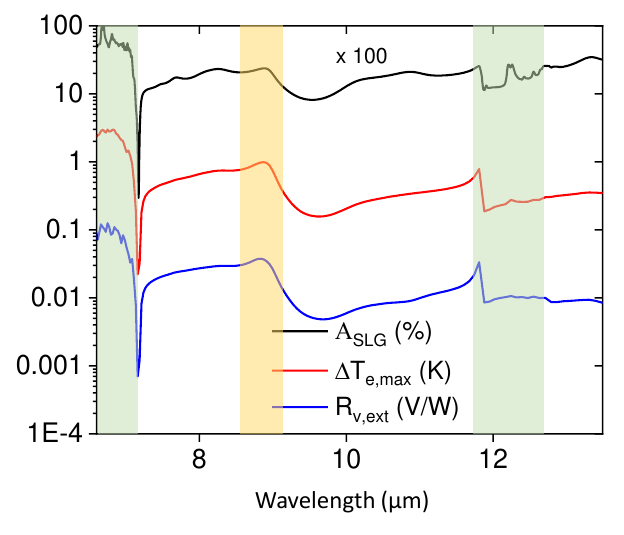}
	\caption{\footnotesize \textbf{a)} Theoretical calculations of the absorption (in percentage), maximum value of the electronic temperature across the graphene channel (in K) and responsivity (in units of V/W) that share the $y$-axis for different units, as mentioned, respectively, which are plotted as a function of the wavelength. We observe that these three variables can be related to each other, as demonstrated in ref. \citenum{Castilla2020}. The highlighted spectral regions correspond to the RBs mentioned in the main text. For the input powers used in our experiments (~1-10 mW and irradiance of 33.6 mW/\textmu m$^2$), the estimated rise in electronic temperature ($\Delta$T$_e$ = T$_{el}$-T$_L$, where T$_L$ is the lattice temperature) is minimal, ranging from $\sim$0.1 K to $\sim$3 K (weak heating regime)\cite{Tielrooij2018, Castilla2019}. Consequently, both the heat capacity and thermal conductivity can be considered temperature-independent and thus absorption-independent, given that the electron temperature rise is linked to absorption.\cite{Castilla2020, Vangelidis2022} Therefore, the cooling length remains unaffected by the spectral dependence of absorption. Our simulations assume a spectrally independent value of 1.5 ps well within this range, resulting in good agreement between the measured and simulated spectral responsivity.\cite{Tielrooij2018, Vangelidis2022} In our study, the carrier density is spectrally independent due to the small electronic temperature as mentioned previously and low photon energy at mid- and long-wave infrared frequencies ($\sim$0.1 eV).
	}
	\label{relation}
\end{figure*}


\begin{figure*}[h!]
	\includegraphics [scale=0.5]
	{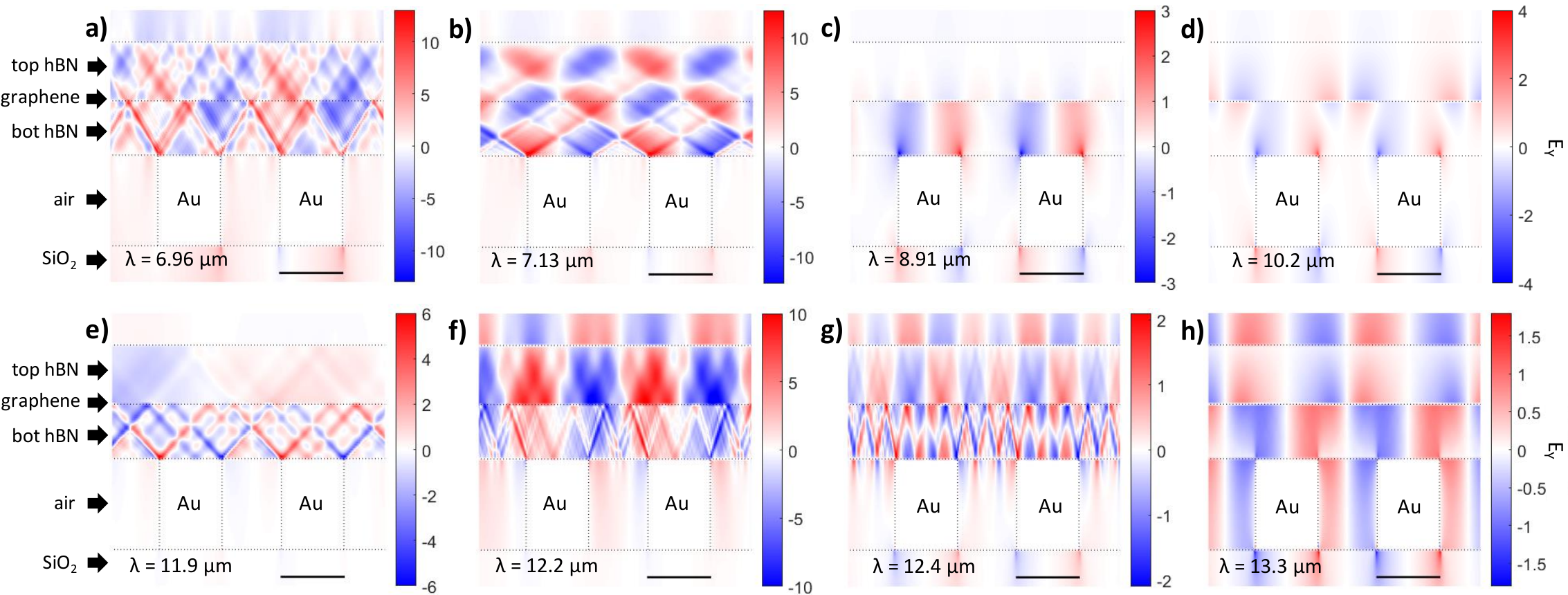}
	\caption{ 
		\footnotesize
		Cross-sectional view of the $y$-component of the electric field normalized to the incident one across a region containing two metal nanorods for illustration. The Z direction (y-axis in graphs) and X direction (x-axis in the graphs) are defined in Fig. 1a in the main text. The black scale bar corresponds to 40 nm. The calculations correspond to a non-uniform graphene Fermi level with a value of 0.4 eV above the metal at \textbf{a)} wavelength 6.96 \textmu m, \textbf{b)} 7.13 \textmu m, \textbf{c)} 8.91 \textmu m, \textbf{d)} 10.2 \textmu m, \textbf{e)} 11.9 \textmu m, \textbf{f)} 12.2 \textmu m, \textbf{g)} 12.4 \textmu m, \textbf{h)} 13.3 \textmu m  corresponding to peak 1, 2, 3, 4, 5, 6, 7 and 8 respectively labelled in Fig. 2a in the main text.
	}
	\label{lwir_sem_pics}
\end{figure*}


\begin{figure*}[h!]
	\includegraphics [scale=0.5]
	{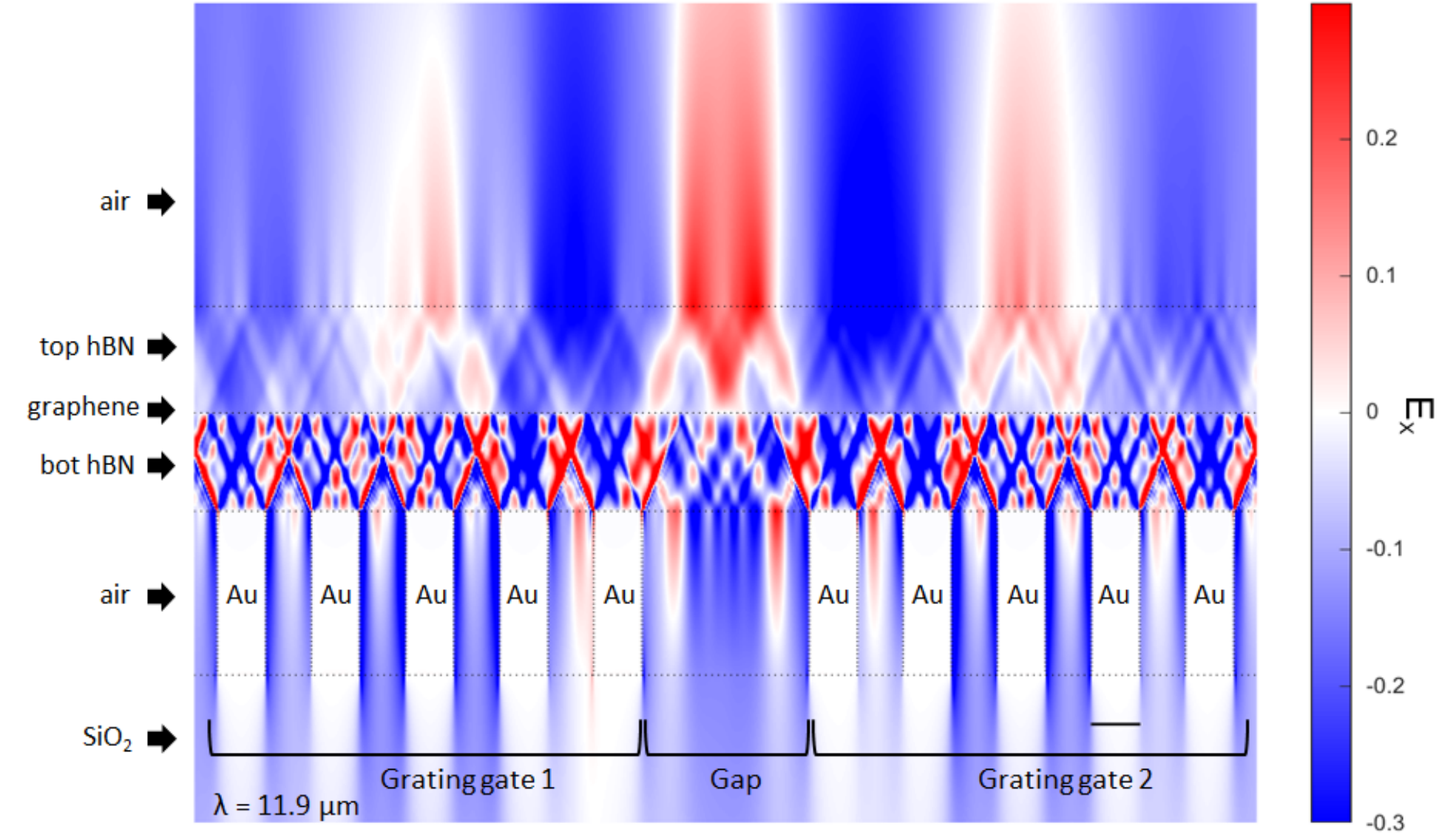}
	\caption{ 
		\footnotesize
		Extended cross-sectional view of the $x-$component of electric field at $\lambda =$ 11.93 \textmu m, including the grating gates 1 and 2 and the gap between them. The black scale bar corresponds to 40 nm. Notably, peak 5 of the Fig. 2 in the main text presents a peculiar resonance by meeting the condition of $k_{\rm eff} = \pi/D$, which is interpreted as a defect mode with the diffraction order $n=1/2$ owing to the broken symmetry of the grating. 
	}
	\label{lwir_sem_pics}
\end{figure*}


\begin{figure*}[h!]
	\includegraphics [scale=0.65]
	{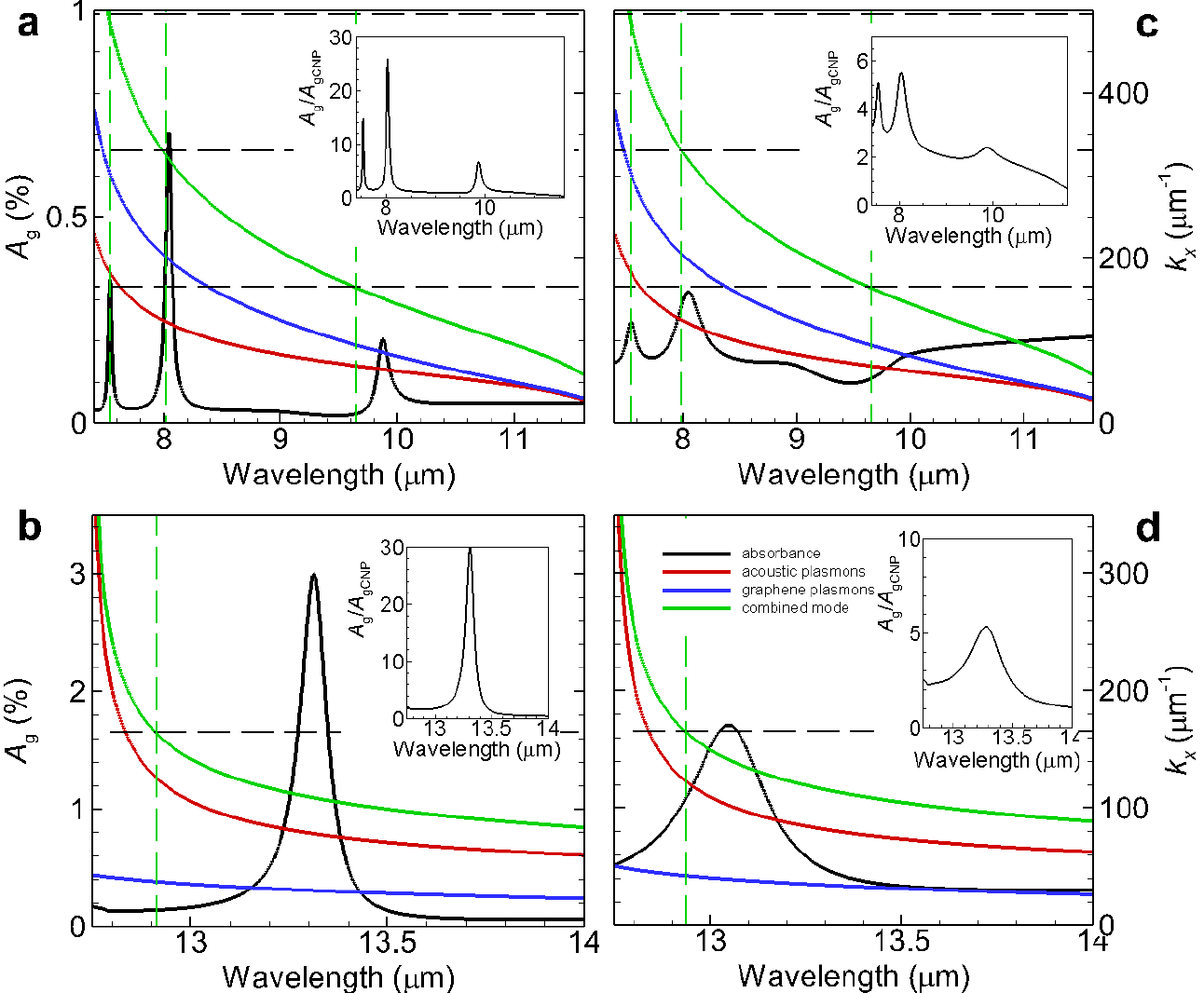}
	\caption{ 
		\footnotesize
		\textbf{a-d)} Theoretically calculated graphene absorbance for the device 2 (black solid lines, values correspond to left vertical axis) and eigenmode wavevector $k_x$ (red, blue and green lines for acoustic plasmons, graphene plasmons and combined modes, respectively, values correspond to right vertical axis) versus vacuum wavelength $\lambda$ (horizontal axis). In each panel \textbf{a-d}, the inset show the graphene absorbance, normalized to the one at the charge neutrality point. All the absorbance are calculated using the Drude model of graphene’s conductivity and for the cases of high relaxation time $\tau$ (panels \textbf{a} and \textbf{b}), defined by Eqs. \ref{eq:sigma-g} and \ref{eq:sigma-gamma} without the $v_F/D$ term (see Supplementary Note 2, section B) and low relaxation time (panels \textbf{c} and \textbf{d}, defined by Eq. \ref{eq:sigma-gamma}). In panels \textbf{a-d}, the lattice vectors $k_x = 4\pi n/D$, with $D$ as the grating period (containing the gap between the metal rods plus the metal width) and ($n=1,2,…$) are depicted by black dashed  horizontal lines (right vertical axis), while their crossing points with combined mode dispersion curves are depicted by green dashed vertical lines.
	}
	\label{lwir_sem_pics}
\end{figure*}


\begin{figure*}[h!]
	\includegraphics [width=\textwidth,scale=0.30]
	{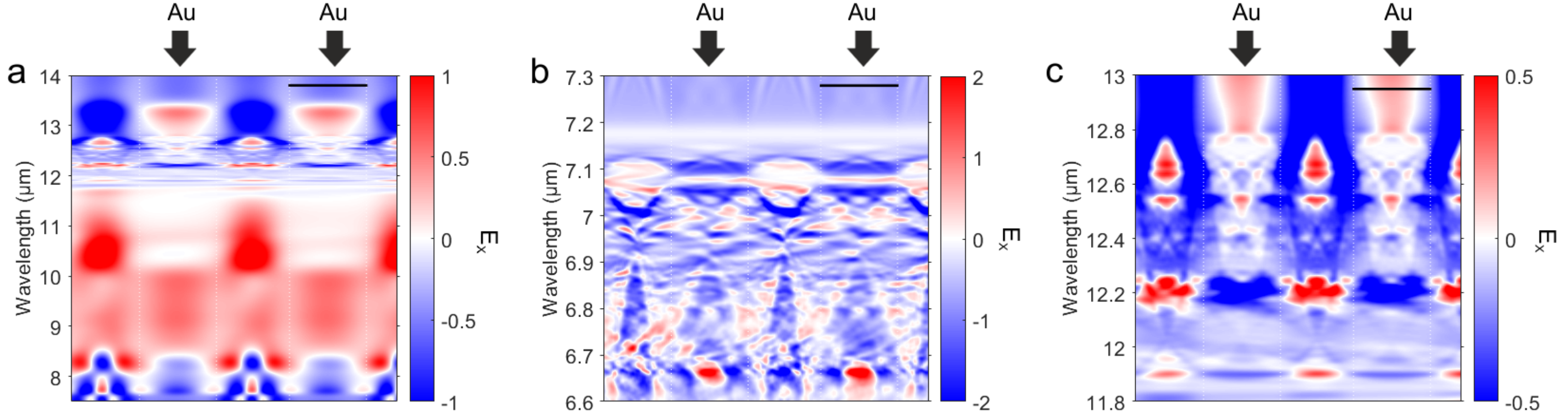}
	
	\caption{ 
		\footnotesize
		\textbf{a)} Electric field intensity ($x-$component) map as a function of the incident wavelength and position along the source-drain direction containing two metal nanorods on the GG2 indicated in black arrows. The Fermi level is 0.41 eV. The black scale bar corresponds to 40 nm.
	}
	\label{lwir_sem_pics}
\end{figure*}


\begin{figure*}[h!]
	\includegraphics [scale=0.65]
	{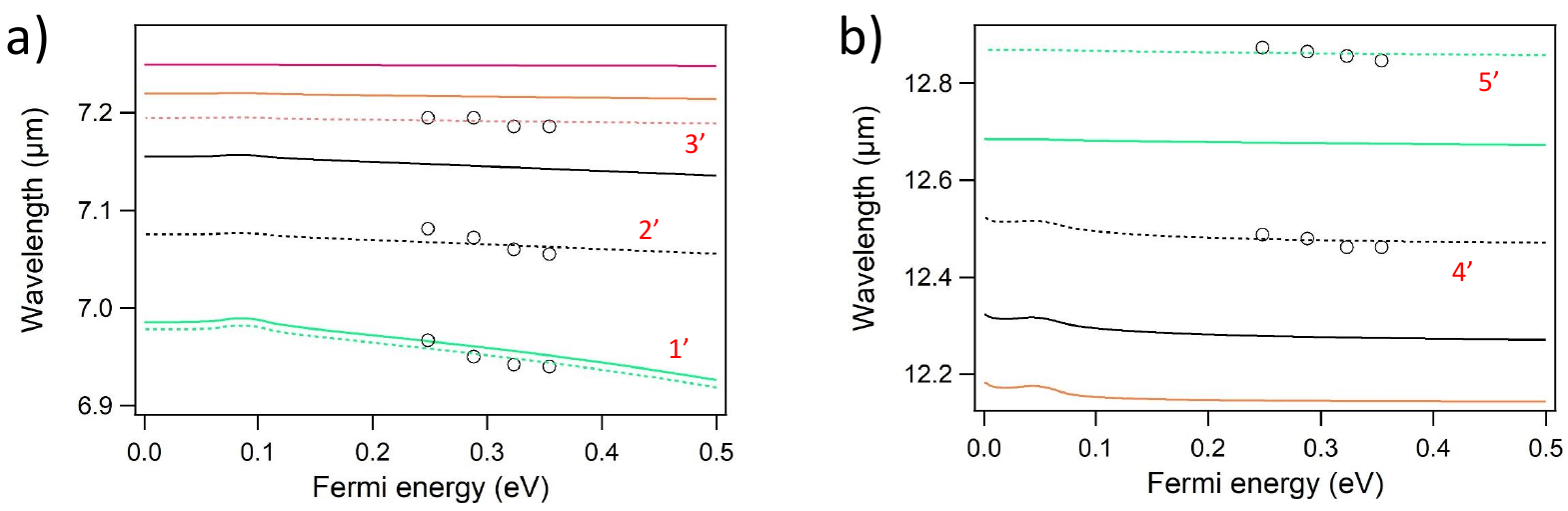}
	\caption{ 
		\footnotesize Hybridized polaritons spectral peak position as a function of the Fermi energy for the \textbf{a)} upper and \textbf{b)} lower hBN RB. The open circles represent the experimental points found in Fig. 3. The solid lines are theoretical calculations and the dashed lines are the same but with a spectral offset to overlap the spectral shift respect to the experimental points. The colors of the lines represent the mode order. We observe that the spectral position of the peaks evolve in a sublinear manner according to $\sqrt{E_F}$, according to peak wavenumber $\propto \sqrt{E_F}^{0.01}$
	}
	\label{lwir_sem_pics}
\end{figure*}


\begin{figure*}[h!]
	\includegraphics[width=\textwidth]
	{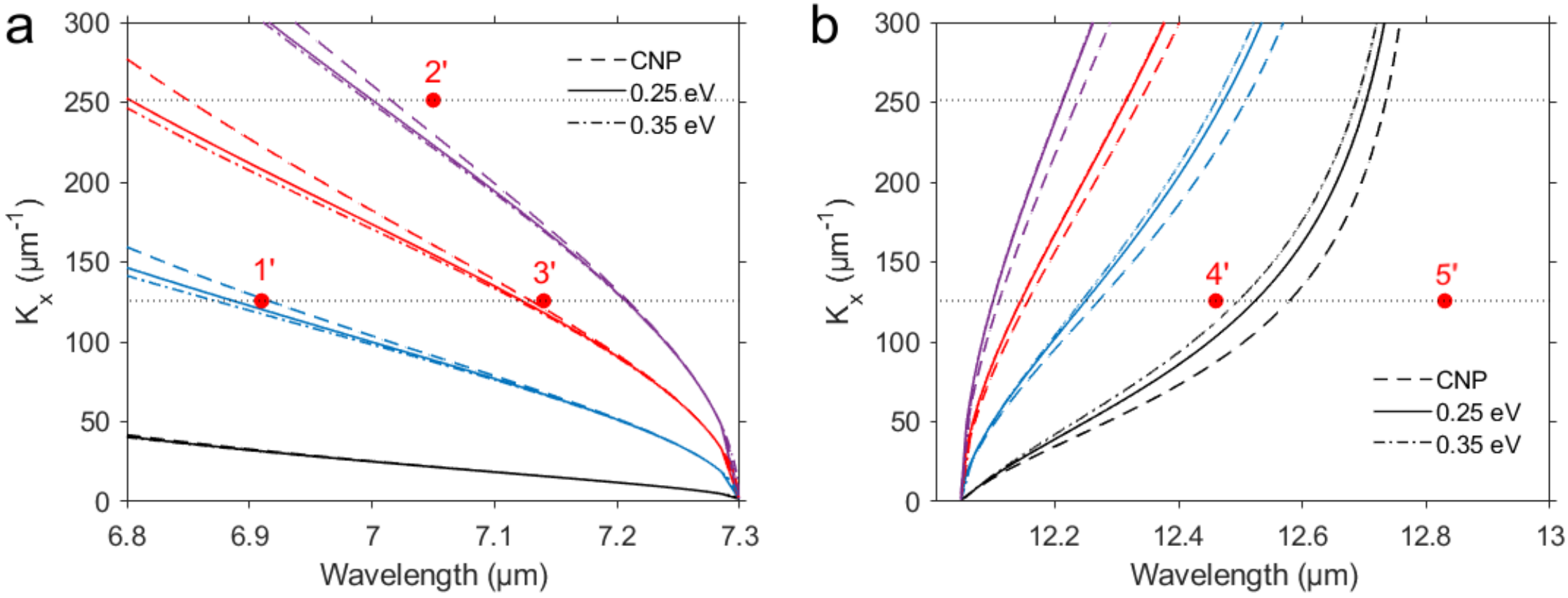}
	\caption{ 
		\footnotesize \textbf{a)} Dispersion relation for device 3 of the hybridized plasmon phonon polariton modes at the upper RB of hBN. The two horizontal dashed lines correspond to the first and second diffraction order resonances launched by the metal rod array. The marked red dots represent the experimental values, which the numeric labels are defined in Fig. 3a in the main text. The graphene Fermi level is 0.35 eV.
		\textbf{b)} Same as panel \textbf{a} but at the lower RB spectral range.
	}
	\label{lwir_sem_pics}
\end{figure*}


\begin{figure*}[h!]
	\includegraphics [scale=0.6]
	{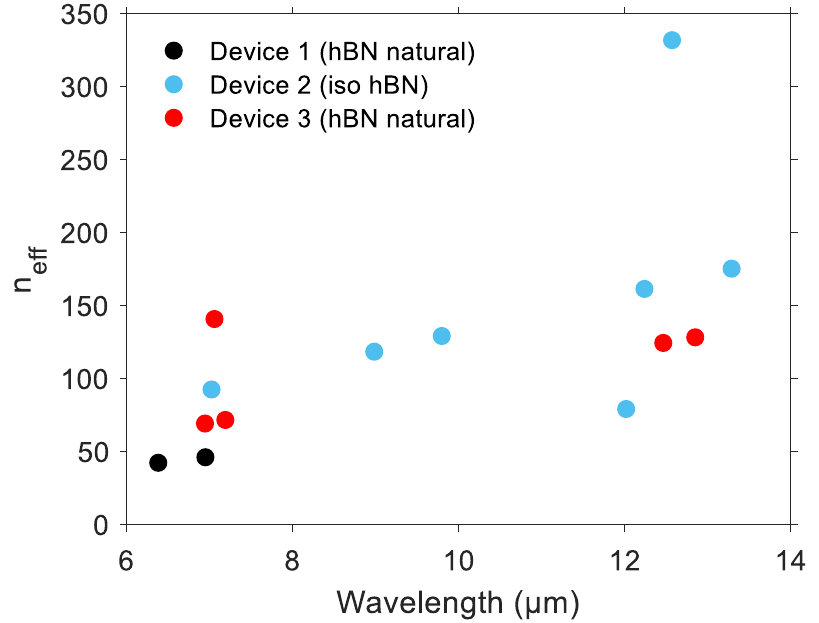}
	\caption{ 
		\footnotesize Effective refractive index (n$_{\rm eff}$) spectrum of the measured 2D polaritonic nanoresonators. The effective refractive index is defined as n$_{\rm eff} \simeq k_{\rm p}/k_{\rm in}$,\cite{Lee2020} where $k_{\rm p} = 2\pi m/period$ is the polariton momentum, $k_{\rm in} = 2\pi/\lambda_{0}$ is the incident wave momentum and $m$ is the order of the resonance determined from the dispersion relations shown in previous graphs.
	}
	\label{lwir_8}
\end{figure*}


\begin{figure*}[h!]
	\includegraphics [scale=0.6]
	{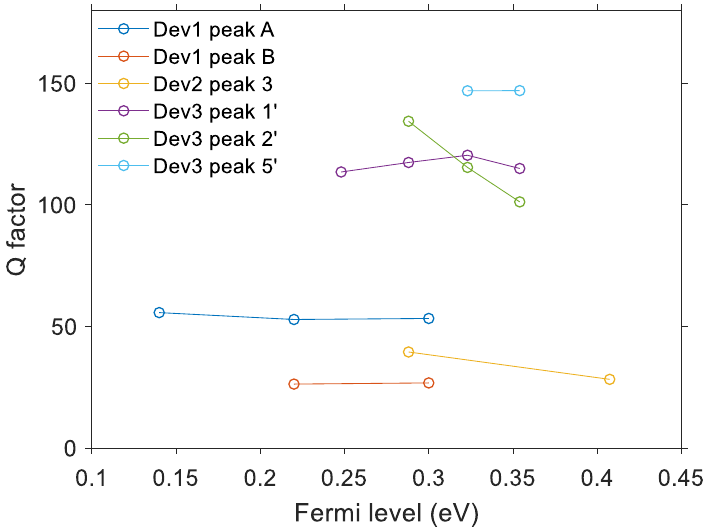}
	\caption{ 
		\footnotesize Q factor values of the peaks observed in the devices 1-3 spectra as a function of the graphene Fermi level. We observe no significant changes of the Q factor when increasing the Fermi level in agreement with previous optical spectroscopy studies\cite{Brar2013c, Brar2014}.
	}
	\label{lwir_8}
\end{figure*}


\clearpage


\maketitle

\section*{Supplementary Note 2: Semi-analytical approach to the modelling of graphene plasmonic crystal.}

\begin{figure}[t]
	\includegraphics[width=10cm]{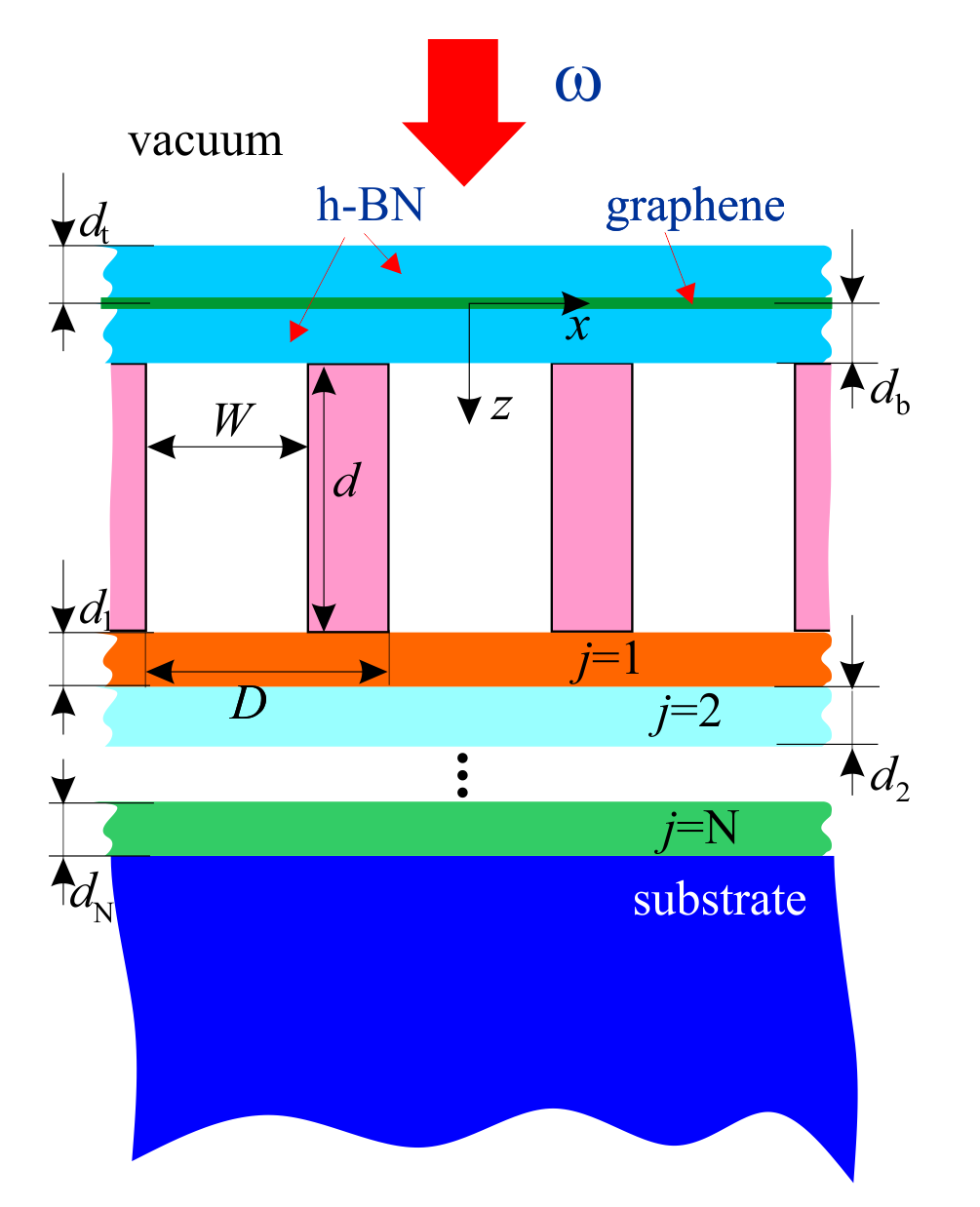}
	
	\caption{Cross-sectional view of the studied geometry: diffraction grating made of PEC, arranged above
		layered substrate, and covered by the graphene layer, encapsulated
		between two hBN slabs. Figure schematic is not to scale.}
	
	\label{fig:geometry}
\end{figure}

We consider the structure, depicted in Fig.\,\ref{fig:geometry}.
The array of perfect electric conductor (PEC) bars is supposed to
be periodic (with period $D$) along the direction of $x$-axis. The
height and width of each PEC bar are $d$ and $D-W$, respectively,
thus each individual PEC bar occupies the spatial domain $d_{b}<z<d_{b}+d$,
$lD+W/2<z<\left(l+1\right)D-W/2$ (here $l$ is the number of period).
From the upper side the array of metal bars is covered by the layered
structure, composed of the hBN bottom layer with thickness $d_{b}$
(at region $0<z<d_{b}$), graphene monolayer (arranged at $z=0$),
and hBN top layer with thickness $d_{t}$ (located at $-d_{t}<z<0$).
Note, that Fermi energy of graphene layer is supposed to be periodic
function of the coordinate $x$, i.e. $E_{F}\left(x\right)=E_{F}\left(x+D\right)$.
This hBN-graphene composite structure is supposed to be truncated
by the semi-infinite vacuum/air, which occupies half-space $z<-d_{t}$.
From bottom side the array of metal bars is deposited on top of another
layered structure, which is composed of $N$ capping layers, where
each individual layer is characterized by thickness $d_{j}$ {[}and
occupies spatial domain $L_{j}<z<L_{j+1},$where $L_{j}=d_{t}+d+\sum_{r=1}^{j-1}d_{r}$
is the coordinate of interface between $\left(j-1\right)$th and $j$th
layers{]} and dielectric permeability $\varepsilon^{(j)}\left(\omega\right)$
{[}here $j=1,...,N${]}. In its turn, this layered structure is deposited
on top of semi-infinite substrate with dielectric function $\varepsilon^{(s)}\left(\omega\right)$,
arranged at $z>L_{N+1}$. We also consider that incident wave with
frequency $\omega$ falls normally on the above structure from vacuum/air
side.

\subsection{Solutions of Maxwell equations}

Assuming electromagnetic field time-dependence as $\mathbf{E},\mathbf{H}\sim\exp{-i\omega t}$,
we represent Maxwell equations for p-polarized wave as
\begin{eqnarray}
\frac{\partial E_{x}^{(\alpha)}}{\partial z}-\frac{\partial E_{z}^{(\alpha)}}{\partial x}=\frac{i\omega}{c}H_{y}^{(\alpha)},\label{eq:max_Hy}\\
-\frac{\partial H_{y}^{(\alpha)}}{\partial z}=-\frac{i\omega}{c}\varepsilon_{xx}^{(\alpha)}E_{x}^{(\alpha)},\label{eq:max_Ex}\\
\frac{\partial H_{y}^{(\alpha)}}{\partial x}=-\frac{i\omega}{c}\varepsilon_{zz}^{(\alpha)}E_{z}^{(\alpha)}.\label{eq:max_Ez}
\end{eqnarray}
where $\omega$ is wave cyclic frequency, $c$ is the velocity of
light in vacuum. The superscripts $\alpha$ correspond to the fields
in different spatial domains, and will be specified below. 

In the vacuum/air region $z<-d_{t}$ the solutions of Maxwell equation
can be represented in the form of Fourier-Floquet series as
\begin{eqnarray}
\left(\begin{array}{c}
H_{y}^{(v)}(x,z)\\
E_{x}^{(v)}(x,z)
\end{array}\right)=\sum_{m=0}^{\infty}\hat{F}_{m}^{(v)}\left(\begin{array}{c}
H_{y}^{(i)}\delta_{m,0}\exp\left[ip_{m}^{(v)}\left(z+d_{t}\right)\right]\\
H_{y||m}^{(r)}\exp\left[-ip_{m}^{(v)}\left(z+d_{t}\right)\right]
\end{array}\right)\times\label{eq:HEv-vacuum}\\
\cos\left[\frac{2\pi m}{D}x\right],\nonumber 
\end{eqnarray}
where 
\begin{eqnarray}
\hat{F}_{m}^{(v)}=\left(\begin{array}{cc}
1 & 1\\
\frac{cp_{m}^{(v)}}{\omega} & -\frac{cp_{m}^{(v)}}{\omega}
\end{array}\right)\label{eq:field-mat-vacuum-m}
\end{eqnarray}
is the field matrix, 
\begin{eqnarray*}
	p_{m}^{(v)}=\sqrt{\left(\frac{\omega}{c}\right)^{2}-\left(\frac{2\pi m}{D}\right)^{2}}
\end{eqnarray*}
is the out-of-plane wavevector component of $m$th harmonics, $H_{y}^{(i)}$
and $H_{y||m}^{(r)}$ are the amplitudes of the magnetic field of
incident and reflected wave of $m$th harmonics. Since for normal
incidence the in-plane wavevector component of incident wave is zero,
the diffracted should be even function along $x$-axis. This fact
allows an expansion with respect to cosine function in Eq.\,(\ref{eq:HEv-vacuum}).
Inside the semi-infinite substrate, $z>L_{N+1}$, the electromagnetic
field can be expressed as
\begin{eqnarray}
\left(\begin{array}{c}
H_{y}^{(s)}(x,z)\\
E_{x}^{(s)}(x,z)
\end{array}\right)=\sum_{m=0}^{\infty}\hat{F}_{m}^{(s)}\left(\begin{array}{c}
H_{y||m}^{(s)}\exp\left[ip_{m}^{(s)}\left(z-L_{N+1}\right)\right]\\
0
\end{array}\right)\times\label{eq:HEs-substrate}\\
\times\cos\left[\frac{2\pi m}{D}x\right].\nonumber 
\end{eqnarray}
Here 
\begin{eqnarray}
\hat{F}_{m}^{(s)}=\left(\begin{array}{cc}
1 & 1\\
\frac{cp_{m}^{(s)}}{\omega\varepsilon^{(s)}\left(\omega\right)} & -\frac{cp_{m}^{(s)}}{\omega\varepsilon^{(s)}\left(\omega\right)}
\end{array}\right),\label{eq:field-mat-substrate-m}
\end{eqnarray}
is the field matrix, $p_{m}^{(s)}=\sqrt{\left(\frac{\omega}{c}\right)^{2}\varepsilon^{(s)}\left(\omega\right)-\left(\frac{2\pi m}{D}\right)^{2}}$
is the out-of-plane component of the wavevector of $m$th harmonics.
In Eq.\,(\ref{eq:HEs-substrate}) zero in the second line means absence
of the backward-propagating wave, and presence of the transmitted
wave only (with amplitude of the magnetic field of $m$th harmonics
$H_{y||m}^{(s)}$). 

Inside the dielectric layers of finite thickness electromagnetic fields
will be represented in the different manner -- in form of the transfer-matrix
\begin{eqnarray}
\hat{Q}_{m}^{(\alpha)}\left(z\right)=\left(\begin{array}{cc}
\cos\left[p_{m}^{(\alpha)}z\right] & \frac{i\omega\varepsilon_{xx}^{(\alpha)}\left(\omega\right)}{cp_{m}^{(\alpha)}}\sin\left[p_{m}^{(\alpha)}z\right]\\
\frac{icp_{m}^{(\alpha)}}{\omega\varepsilon_{xx}^{(\alpha)}\left(\omega\right)}\sin\left[p_{m}^{(\alpha)}z\right] & \cos\left[p_{m}^{(\alpha)}z\right]
\end{array}\right),\label{eq:Qm}
\end{eqnarray}
which is written for the general case of anisotropic medium. For this
case the out-of-plane component of wavevector is represented as$p_{m}^{(\alpha)}=\sqrt{\left(\frac{\omega}{c}\right)^{2}\varepsilon_{xx}^{(\alpha)}\left(\omega\right)-\left(\frac{2\pi m}{D}\right)^{2}\varepsilon_{xx}^{(\alpha)}\left(\omega\right)/\varepsilon_{zz}^{(\alpha)}\left(\omega\right)}$.
To be more specific, total field in different layers of structures
will be represented in the following manner, 
\begin{eqnarray}
\left(\begin{array}{c}
H_{y}^{(t)}(x,z)\\
E_{x}^{(t)}(x,z)
\end{array}\right)=\sum_{m=0}^{\infty}\hat{Q}_{m}^{(t)}\left(z+d_{t}\right)\times\nonumber \\
\times\left(\begin{array}{c}
h_{y||m}^{(t)}\left(-d_{t}\right)\\
e_{x||m}^{(t)}\left(-d_{t}\right)
\end{array}\right)\cos\left[\frac{2\pi m}{D}x\right]\label{eq:HE-top-hBN}
\end{eqnarray}
inside top hBN layer at $-d_{t}<z<0$, 
\begin{eqnarray}
\left(\begin{array}{c}
H_{y}^{(b)}(x,z)\\
E_{x}^{(b)}(x,z)
\end{array}\right)=\sum_{m=0}^{\infty}\hat{Q}_{m}^{(b)}\left(z\right)\times\nonumber \\
\times\left(\begin{array}{c}
h_{y||m}^{(b)}\left(0\right)\\
e_{x||m}^{(b)}\left(0\right)
\end{array}\right)\cos\left[\frac{2\pi m}{D}x\right]\label{eq:HE-bottom-hBN}
\end{eqnarray}
inside bottom hBN layer at $0<z<d_{b}$, and 
\begin{eqnarray}
\left(\begin{array}{c}
H_{y}^{(j)}(x,z)\\
E_{x}^{(j)}(x,z)
\end{array}\right)=\sum_{m=0}^{\infty}\hat{Q}_{m}^{(j)}\left(z-L_{j+1}\right)\times\nonumber \\
\times\left(\begin{array}{c}
h_{y||m}^{(j)}\left(L_{j+1}\right)\\
e_{x||m}^{(j)}\left(L_{j+1}\right)
\end{array}\right)\cos\left[\frac{2\pi m}{D}x\right]\label{eq:HE-j}
\end{eqnarray}
for $j$th layer in the bottom layered structure $L_{j}<z<L_{j+1}$.
In the medium $d_{b}<z<d_{b}+d$, occupied by the PEC bars, the electromagnetic
field can be represented as the superposition of waveguide modes inside
the gaps. Thus, inside the spatial domain $lD-W/2<x<lD+W/2$ (slits
of width $W$) the tangential components of the electromagnetic field
can be written as 
\begin{eqnarray}
\left(\begin{array}{c}
H_{y||l}^{(g)}(x,z)\\
E_{x||l}^{(g)}(x,z)
\end{array}\right)=iW\sum_{n=0}^{\infty}\cos\left[\frac{2n\pi}{W}\left(x+\frac{W}{2}-lL\right)\right]\times\label{eq:HE-g}\\
\left(\begin{array}{cc}
\omega/c & \omega/c\\
\nu_{n} & -\nu_{n}
\end{array}\right)\left(\begin{array}{c}
A_{n}^{(+,l)}\exp\left[i\nu_{n}\left(z-d_{b}\right)\right]\\
A_{n}^{(-,l)}\exp\left[-i\nu_{n}\left(z-d_{b}\right)\right]
\end{array}\right),\nonumber 
\end{eqnarray}
where $\nu_{n}=\sqrt{\left(\frac{\omega}{c}\right)^{2}-\left(\frac{2n\pi}{W}\right)^{2}}$
, $A_{n}^{(\pm,l)}$ are the amplitudes of forward- and backward-propagating
waves of the $n$th eigenmode in the $l$th gap. 

\subsection{Equation of the amplitudes of waveguide modes}

As the boundary conditions at interfaces $z=-d_{t}$, and $z=L_{j}$
($j=2,...,N+1$) between homogeneous (in $x$-direction) dielectric
layers without graphene we use continuity of tangential components
of electromagnetic waves. At the same time at interface $z=0$, containing
graphene, tangential component of the electric field is also continuous
across the interface, while the magnetic field tangential component
is discontinuous across the interface owing the the presence of 2D
currents in graphene $H_{y}^{(b)}\left(x,0\right)-H_{y}^{(t)}\left(x,0\right)=-\left(4\pi/c\right)\sigma^{(g)}\left(\omega,x\right)E_{x}^{(t)}\left(x,0\right)$,
where 
\begin{eqnarray}
\sigma^{(g)}\left(\omega,x\right)=\frac{e^{2}}{\hbar^{2}\pi}\frac{E_{F}\left(x\right)}{\gamma\left(x\right)-i\omega}\label{eq:sigma-g}
\end{eqnarray}
is the Drude conductivity of graphene. The inverse relaxation time
$\gamma\left(x\right)$ is also considered to be coordinate-dependent
{[}as well as Fermi energy $E_{F}\left(x\right)${]}. Being expressed
through the electron mobility in graphene $\mu$ (which is considered
to be constant), the inverse relaxation can be defined as 
\begin{eqnarray}
\gamma\left(x\right)=ev_{F}^{2}/\left[\mu E_{F}\left(x\right)\right]+v_{F}/D.\label{eq:sigma-gamma}
\end{eqnarray}
At surfaces of periodic grating, $z=d_{b}+d$ and $z=d_{b}+d$, the
medium is non-homogeneous, here the imposed boundary conditions are
of more complicated form: the continuity of electric and magnetic
field tangential components at gaps regions (between PEC bars, $lD-W/2<x<lD+W/2$),
and nullity of the electric field tangential component at the surfaces
of PEC bars, $lD+W/2<x<\left(l+1\right)D-W/2$.

Applying these boundary conditions, as well as Bloch theorem for normally
incident wave $A_{n}^{(\pm,0)}\equiv A_{n}^{(\pm,l)}$, it is possible
to obtain the system of coupled linear equations, which governs the
amplitudes of the waveguide modes, 
\begin{eqnarray}
i\frac{\omega W}{c}\frac{1+\delta_{n^{\prime},0}}{2}\left\{ A_{n^{\prime}}^{(+,0)}+A_{n^{\prime}}^{(-,0)}\right\} -\nonumber \\
i\frac{W^{2}}{D}\sum_{n=0}^{\infty}\nu_{n}\left\{ A_{n}^{(+,0)}-A_{n}^{(-,0)}\right\} \left[\mathcal{P}_{n^{\prime}}\right]^{T}\hat{\mathcal{F}}_{12}^{(v,tot)}\left[\hat{\mathcal{F}}_{22}^{(v,tot)}\right]^{-1}\hat{\mathcal{N}}\mathcal{P}_{n}=\label{eq:Apm-eq1}\\
\left[\mathcal{P}_{n^{\prime}}\right]^{T}\left\{ \hat{\mathcal{F}}_{11}^{(v,tot)}-\hat{\mathcal{F}}_{12}^{(v,tot)}\left[\hat{\mathcal{F}}_{22}^{(v,tot)}\right]^{-1}\hat{\mathcal{F}}_{21}^{(v,tot)}\right\} \mathcal{H}_{y}^{(i)},\nonumber \\
i\frac{\omega W}{c}\frac{1+\delta_{n^{\prime},0}}{2}\left\{ A_{n^{\prime}}^{(+,0)}\exp\left(i\nu_{n^{\prime}}d\right)+A_{n^{\prime}}^{(-,0)}\exp\left(-i\nu_{n^{\prime}}d\right)\right\} -\label{eq:Apm-eq2}\\
i\frac{W^{2}}{D}\sum_{n=0}^{\infty}\nu_{n}\left\{ A_{n}^{(+,0)}\exp\left(i\nu_{n}d\right)-A_{n}^{(-,0)}\exp\left(-i\nu_{n}d\right)\right\} \left[\mathcal{P}_{n^{\prime}}\right]^{T}\hat{\mathcal{F}}_{11}^{(s,tot)}\left[\hat{\mathcal{F}}_{21}^{(s,tot)}\right]^{-1}\hat{\mathcal{N}}\mathcal{P}_{n}=0.\nonumber 
\end{eqnarray}
Here $\mathcal{P}_{n}=\left(P_{n||0},\thinspace P_{n||1},\thinspace...\right)^{T}$
is the column vector, whose elements are 
\begin{eqnarray*}
	P_{n||m}=\frac{DW}{\pi}\frac{m}{\left(mW\right)^{2}-\left(nD\right)^{2}}\sin\left(\frac{\pi mW}{D}\right),
\end{eqnarray*}
$\hat{\mathcal{N}}$ is the diagonal matrix with elements $\hat{\mathcal{N}}_{m^{\prime},m}=\left(2-\delta_{m,0}\right)\delta_{m,m^{\prime}}$,
$\mathcal{H}_{y}^{(i)}=\left(H_{y}^{(i)},0,0,...\right)^{T}$ is the
incident wave's column vector. Also two total field matrices,
\begin{eqnarray}
\hat{\mathcal{F}}^{(v,tot)}=\left[\hat{\mathcal{Q}}^{(b)}\left(d_{b}\right)\right]\hat{\mathcal{Q}}^{(g)}\left[\hat{\mathcal{Q}}^{(t)}\left(d_{t}\right)\right]\hat{\mathcal{F}}^{(v)}=\left(\begin{array}{cc}
\hat{\mathcal{F}}_{11}^{(v,tot)} & \hat{\mathcal{F}}_{12}^{(v,tot)}\\
\hat{\mathcal{F}}_{21}^{(v,tot)} & \hat{\mathcal{F}}_{22}^{(v,tot)}
\end{array}\right),\label{eq:Fvtot}\\
\hat{\mathcal{F}}^{(s,tot)}=\left[\prod_{j=1}^{N}\hat{\mathcal{Q}}^{(j)}\left(-d_{j}\right)\right]\hat{\mathcal{F}}^{(s)}=\left(\begin{array}{cc}
\hat{\mathcal{F}}_{11}^{(s,tot)} & \hat{\mathcal{F}}_{12}^{(s,tot)}\\
\hat{\mathcal{F}}_{21}^{(s,tot)} & \hat{\mathcal{F}}_{22}^{(s,tot)}
\end{array}\right),\label{eq:Fstot}
\end{eqnarray}
are block matrices, composed of four submatrices and are obtained
by multiplication of a series of other block matrices. Among them
matrices
\begin{eqnarray*}
	\hat{\mathcal{F}}^{(s)}=\left(\begin{array}{cc}
		\hat{\mathcal{F}}_{11}^{(s)} & \hat{\mathcal{F}}_{12}^{(s)}\\
		\hat{\mathcal{F}}_{21}^{(s)} & \hat{\mathcal{F}}_{22}^{(s)}
	\end{array}\right),\\
	\hat{\mathcal{F}}^{(v)}=\left(\begin{array}{cc}
		\hat{\mathcal{F}}_{11}^{(v)} & \hat{\mathcal{F}}_{12}^{(v)}\\
		\hat{\mathcal{F}}_{21}^{(v)} & \hat{\mathcal{F}}_{22}^{(v)}
	\end{array}\right),\\
	\hat{\mathcal{Q}}^{(j)}\left(-d_{j}\right)=\left(\begin{array}{cc}
		\hat{\mathcal{Q}}_{11}^{(j)}\left(-d_{j}\right) & \hat{\mathcal{Q}}_{12}^{(j)}\left(-d_{j}\right)\\
		\hat{\mathcal{Q}}_{21}^{(j)}\left(-d_{j}\right) & \hat{\mathcal{Q}}_{22}^{(j)}\left(-d_{j}\right)
	\end{array}\right),\\
	\hat{\mathcal{Q}}^{(\alpha)}\left(d_{\alpha}\right)=\left(\begin{array}{cc}
		\hat{\mathcal{Q}}_{11}^{(\alpha)}\left(d_{\alpha}\right) & \hat{\mathcal{Q}}_{12}^{(\alpha)}\left(d_{\alpha}\right)\\
		\hat{\mathcal{Q}}_{21}^{(\alpha)}\left(d_{\alpha}\right) & \hat{\mathcal{Q}}_{22}^{(\alpha)}\left(d_{\alpha}\right)
	\end{array}\right),\quad\alpha=t,b
\end{eqnarray*}
are characterized by the fact that their submatrices are diagonal
with elements 
\begin{eqnarray*}
	\hat{\mathcal{F}}_{11}^{(v)}=\hat{\mathcal{F}}_{12}^{(v)}=\hat{\mathcal{F}}_{11}^{(s)}=\hat{\mathcal{F}}_{12}^{(s)}=\mathcal{\hat{I}},\\
	\left[\hat{\mathcal{F}}_{21}^{(v)}\right]_{m^{\prime},m}=-\left[\hat{\mathcal{F}}_{22}^{(v)}\right]_{m^{\prime},m}=\frac{cp_{m}^{(v)}}{\omega}\delta_{m^{\prime},m},\\
	\left[\hat{\mathcal{F}}_{21}^{(s)}\right]_{m^{\prime},m}=-\left[\hat{\mathcal{F}}_{22}^{(s)}\right]_{m^{\prime},m}=\frac{cp_{m}^{(s)}}{\omega\varepsilon^{(s)}\left(\omega\right)}\delta_{m^{\prime},m},\\
	\left[\hat{\mathcal{Q}}_{11}^{(j)}\left(-d_{j}\right)\right]_{m^{\prime},m}=\cos\left[p_{m}^{(j)}d_{j}\right]\delta_{m^{\prime},m},\\
	\left[\hat{\mathcal{Q}}_{12}^{(j)}\left(-d_{j}\right)\right]_{m^{\prime},m}=-\frac{i\omega\varepsilon^{(j)}\left(\omega\right)}{cp_{m}^{(j)}}\sin\left[p_{m}^{(j)}d_{j}\right]\delta_{m^{\prime},m},\\
	\left[\hat{\mathcal{Q}}_{21}^{(j)}\left(-d_{j}\right)\right]_{m^{\prime},m}=-\frac{icp_{m}^{(j)}}{\omega\varepsilon^{(j)}\left(\omega\right)}\sin\left[p_{m}^{(j)}d_{j}\right]\delta_{m^{\prime},m},\\
	\left[\hat{\mathcal{Q}}_{22}^{(j)}\left(-d_{j}\right)\right]_{m^{\prime},m}=\cos\left[p_{m}^{(j)}d_{j}\right]\delta_{m^{\prime},m},\\
	\left[\hat{\mathcal{Q}}_{11}^{(\alpha)}\left(-d_{\alpha}\right)\right]_{m^{\prime},m}=\cos\left[p_{m}^{(\alpha)}d_{\alpha}\right]\delta_{m^{\prime},m},\\
	\left[\hat{\mathcal{Q}}_{12}^{(\alpha)}\left(d_{\alpha}\right)\right]_{m^{\prime},m}=\frac{i\omega\varepsilon_{xx}^{(\alpha)}\left(\omega\right)}{cp_{m}^{(\alpha)}}\sin\left[p_{m}^{(\alpha)}d_{\alpha}\right]\delta_{m^{\prime},m},\\
	\left[\hat{\mathcal{Q}}_{21}^{(\alpha)}\left(d_{\alpha}\right)\right]_{m^{\prime},m}=\frac{icp_{m}^{(\alpha)}}{\omega\varepsilon_{xx}^{(\alpha)}\left(\omega\right)}\sin\left[p_{m}^{(\alpha)}d_{\alpha}\right]\delta_{m^{\prime},m},\\
	\left[\hat{\mathcal{Q}}_{22}^{(\alpha)}\left(d_{\alpha}\right)\right]_{m^{\prime},m}=\cos\left[p_{m}^{(\alpha)}d_{\alpha}\right]\delta_{m^{\prime},m},
\end{eqnarray*}
$\mathcal{\hat{I}}$ is the diagonal unity matrix. At the same time
matrix
\begin{eqnarray*}
	\hat{\mathcal{Q}}^{(g)}=\left(\begin{array}{cc}
		\mathcal{\hat{I}} & -\frac{4\pi}{c}\hat{\Sigma}\\
		0 & \mathcal{\hat{I}}
	\end{array}\right),
\end{eqnarray*}
possesses one non-diagonal submatrix $\hat{\Sigma}$, which appears
owing to the nonuniformity of graphene's conductivity across one period
of the structure. The elements of this submatrix are 
\begin{eqnarray*}
	\left[\hat{\Sigma}\right]_{m^{\prime},m}=\frac{1}{2}\left[\frac{1+\delta_{m^{\prime}+m,0}}{1+\delta_{m^{\prime},0}}\sigma_{m+m^{\prime}}^{(g)}\left(\omega\right)+\frac{1+\delta_{m^{\prime}-m,0}}{1+\delta_{m^{\prime},0}}\sigma_{\left|m-m^{\prime}\right|}^{(g)}\left(\omega\right)\right],
\end{eqnarray*}
where 
\begin{eqnarray*}
	\sigma_{m}^{(g)}\left(\omega\right) & =\frac{4}{D\left(1+\delta_{m,0}\right)}\intop_{0}^{D/2}dx\thinspace\sigma^{(g)}\left(\omega,x\right)\cos\left(\frac{2\pi m}{D}x\right)
\end{eqnarray*}
is the Fourier component of graphene's conductivity such that 
\begin{eqnarray*}
	\sigma^{(g)}\left(\omega,x\right)=\sum_{m=0}^{\infty}\sigma_{m}^{(g)}\left(\omega\right)\cos\left(\frac{2\pi m}{D}x\right).
\end{eqnarray*}
After solving Eqs.(\ref{eq:Apm-eq1}) and (\ref{eq:Apm-eq2}), from
amplitudes $A_{n}^{(\pm,0)}$ it is possible to obtain amplitudes
of harmonics of reflected, $\mathcal{H}_{y}^{(r)}=\left(H_{y||0}^{(r)},H_{y||1}^{(r)},H_{y||2}^{(r)},...\right)^{T}$,
and transmitted, $\mathcal{H}_{y}^{(s)}=\left(H_{y||0}^{(s)},H_{y||1}^{(s)},H_{y||2}^{(s)},...\right)$,
waves in the form
\begin{eqnarray}
\mathcal{H}_{y}^{(r)}=i\frac{W^{2}}{D}\sum_{n=0}^{\infty}\nu_{n}\left\{ A_{n}^{(+,l)}-A_{n}^{(-,l)}\right\} \left[\hat{\mathcal{F}}_{22}^{(v,tot)}\right]^{-1}\hat{\mathcal{N}}\mathcal{P}_{n}-\left[\hat{\mathcal{F}}_{22}^{(v,tot)}\right]^{-1}\hat{\mathcal{F}}_{21}^{(v,tot)}\mathcal{H}_{y}^{(i)}.\label{eq:hyr-1}\\
\mathcal{H}_{y}^{(s)}=i\frac{W^{2}}{D}\sum_{n=0}^{\infty}\nu_{n}\left\{ A_{n}^{(+,l)}\exp\left(i\nu_{n}d\right)-A_{n}^{(-,l)}\exp\left(-i\nu_{n}d\right)\right\} \left[\hat{\mathcal{F}}_{21}^{(s,tot)}\right]^{-1}\hat{\mathcal{N}}\mathcal{P}_{n}.
\end{eqnarray}

\subsection{Reflectance, transmittance and absorbance}

The reflectance (transmittance) coefficients can be obtained as the
ratio between $z$-component of Poynting vector of reflected (transmitted)
wave's relation and that of incident wave,
\begin{eqnarray}
R=-\frac{\mathrm{Re}\left\{ \left[\mathcal{H}_{y}^{(r)}\right]^{\dagger}\hat{\mathcal{F}}_{22}^{(v)}\mathcal{\hat{N}}^{-1}\mathcal{H}_{y}^{(r)}\right\} }{\left|H_{y}^{(i)}\right|^{2}},\label{eq:reflectance}\\
T=\frac{\mathrm{Re}\left\{ \left[\mathcal{H}_{y}^{(s)}\right]^{\dagger}\hat{\mathcal{F}}_{21}^{(s)}\mathcal{\hat{N}}^{-1}\mathcal{H}_{y}^{(s)}\right\} }{\left|H_{y}^{(i)}\right|^{2}}.\label{eq:transmittance}
\end{eqnarray}
In Eq.\,(\ref{eq:reflectance}) for reflection coefficient sign minus
appears owing the the propagation of reflected wave in negative direction
of axis $z$, while in Eq.\,(\ref{eq:transmittance}) for transmittance
coefficient sign plus is accounted to the fact, that transmitted wave
propagates in the positive direction of $z$-axis. The total absorption
$A$ of the structure can be calculated as 
\begin{eqnarray*}
	A=1-R-T,
\end{eqnarray*}
while absorption by the graphene only, $A_{g}$, can be represented
as
\begin{eqnarray*}
	A_{g}=\frac{\mathrm{Re}\left\{ \left[\mathcal{E}_{x}^{(t)}\left(0\right)\right]^{\dagger}\mathcal{\hat{N}}^{-1}\hat{\Sigma}\mathcal{E}_{x}^{(t)}\left(0\right)\right\} }{\left|H_{y}^{(i)}\right|^{2}},
\end{eqnarray*}
where $\mathcal{E}_{x}^{(t)}\left(0\right)=\left(e_{x||0}^{(t)}\left(0\right),\thinspace e_{x||1}^{(t)}\left(0\right),\thinspace...\right)^{T}$
is the column vector of electric field harmonic amplitudes on the
graphene.

\section{Phonon-plasmon-polaritons eigenmodes.}

The dispersion relation of the phonon-plasmon-polaritons in the graphene-hBN
hybrid structure can be calculated via transfer-matrix method. For
the eigenmode problem field matrices in the vacuum and substrate can
be obtained by formal substitution of $2\pi m/D=k_{x}$ into respective
field matrices for the excitation problem {[}see Eqs.\,(\ref{eq:transmittance})
and (\ref{eq:field-mat-substrate-m}){]}, thus obtaining 
\begin{eqnarray}
\hat{F}_{k_{x}}^{(v)}=\left(\begin{array}{cc}
1 & 1\\
\frac{cp_{k_{x}}^{(v)}}{\omega} & -\frac{cp_{k_{x}}^{(v)}}{\omega}
\end{array}\right)\label{eq:field-mat-vacuum-kx}\\
\hat{F}_{k_{x}}^{(s)}=\left(\begin{array}{cc}
1 & 1\\
\frac{cp_{k_{x}}^{(s)}}{\omega\varepsilon^{(s)}\left(\omega\right)} & -\frac{cp_{k_{x}}^{(s)}}{\omega\varepsilon^{(s)}\left(\omega\right)}
\end{array}\right),\label{eq:field-mat-substrate-kx}
\end{eqnarray}
where $p_{k_{x}}^{(v)}=\sqrt{\left(\frac{\omega}{c}\right)^{2}-k_{x}^{2}}$,
$p_{k_{x}}^{(s)}=\sqrt{\left(\frac{\omega}{c}\right)^{2}\varepsilon^{(s)}\left(\omega\right)-k_{x}^{2}}$
are out-of-plane component of wavevectors inside the vacuum and substrate,
respectively. Using field matrices \ref{eq:field-mat-vacuum-kx}and
\ref{eq:field-mat-substrate-kx}, fields at the hBN-vacuum and substrate
interfaces can be represented as
\begin{eqnarray}
\left(\begin{array}{c}
h_{y||k_{x}}^{(v)}\left(-d_{t}\right)\\
e_{x||k_{x}}^{(v)}\left(-d_{t}\right)
\end{array}\right)=\hat{F}_{k_{x}}^{(v)}\left(\begin{array}{c}
0\\
H_{y||k_{x}}^{(r)}
\end{array}\right),\label{eq:HE-vacuum-kx}\\
\left(\begin{array}{c}
h_{y||k_{x}}^{(s)}\left(L_{N+1}\right)\\
e_{x||k_{x}}^{(s)}\left(L_{N+1}\right)
\end{array}\right)=\hat{F}_{k_{x}}^{(s)}\left(\begin{array}{c}
H_{y||k_{x}}^{(s)}\\
0
\end{array}\right).\label{eq:HEs-substrate-kx}
\end{eqnarray}
Zero at the first row of Eq.\,\ref{eq:HE-vacuum-kx} describes absence
of incident wave, propagating in the positive direction of $z$-axis
-- a situation, typical for eigenmodes {[}compare with Eq.\,\ref{eq:HEv-vacuum}{]},
while zero in second row of Eq.\,\ref{eq:HEs-substrate-kx} has the
same sense, as that in Eq.\,\ref{eq:field-mat-substrate-m}. At the
same time fields at opposite sides of finite-thickness layers can
be related via transfer-matrices, namely 
\begin{eqnarray}
\left(\begin{array}{c}
h_{y||k_{x}}^{(t)}\left(0\right)\\
e_{x||k_{x}}^{(t)}\left(0\right)
\end{array}\right)=\hat{Q}_{k_{x}}^{(t)}\left(d_{t}\right)\left(\begin{array}{c}
h_{y||k_{x}}^{(t)}\left(-d_{t}\right)\\
e_{x||k_{x}}^{(t)}\left(-d_{t}\right)
\end{array}\right)\label{eq:HE-top-hBN-kx}
\end{eqnarray}
inside top hBN layer at $-d_{t}<z<0$, 
\begin{eqnarray}
\left(\begin{array}{c}
h_{y||k_{x}}^{(b)}\left(d_{b}\right)\\
e_{x||k_{x}}^{(b)}\left(d_{b}\right)
\end{array}\right)=\hat{Q}_{k_{x}}^{(b)}\left(d_{b}\right)\left(\begin{array}{c}
h_{y||k_{x}}^{(b)}\left(0\right)\\
e_{x||k_{x}}^{(b)}\left(0\right)
\end{array}\right)\label{eq:HE-bottom-hBN-kx}
\end{eqnarray}
inside bottom hBN layer at $0<z<d_{b}$, and 
\begin{eqnarray}
\left(\begin{array}{c}
h_{y||k_{x}}^{(j)}\left(L_{j}\right)\\
e_{x||k_{x}}^{(j)}\left(L_{j}\right)
\end{array}\right)=\hat{Q}_{k_{x}}^{(j)}\left(-d_{j}\right)\left(\begin{array}{c}
h_{y||k_{x}}^{(j)}\left(L_{j+1}\right)\\
e_{x||k_{x}}^{(j)}\left(L_{j+1}\right)
\end{array}\right)\label{eq:HE-j-kx}
\end{eqnarray}
for $j$th layer in the bottom layered atructure $L_{j}<z<L_{j+1}$.
In the above equations the transfer-matrices are obtained from Eq.\,\ref{eq:Qm}
by the substitution of $2\pi m/D=k_{x}$,
\begin{eqnarray}
\hat{Q}_{k_{x}}^{(\alpha)}\left(z\right)=\left(\begin{array}{cc}
\cos\left[p_{k_{x}}^{(\alpha)}z\right] & \frac{i\omega\varepsilon_{xx}^{(\alpha)}\left(\omega\right)}{cp_{k_{x}}^{(\alpha)}}\sin\left[p_{k_{x}}^{(\alpha)}z\right]\\
\frac{icp_{k_{x}}^{(\alpha)}}{\omega\varepsilon_{xx}^{(\alpha)}\left(\omega\right)}\sin\left[p_{k_{x}}^{(\alpha)}z\right] & \cos\left[p_{k_{x}}^{(\alpha)}z\right]
\end{array}\right),\label{eq:Qm-kx}\\
p_{k_{x}}^{(\alpha)}=\sqrt{\left(\frac{\omega}{c}\right)^{2}\varepsilon_{xx}^{(\alpha)}\left(\omega\right)-k_{x}^{2}\varepsilon_{xx}^{(\alpha)}\left(\omega\right)/\varepsilon_{zz}^{(\alpha)}\left(\omega\right)}.\nonumber 
\end{eqnarray}
Boundary conditions at interfaces without graphene are continuity
of electric and magnetic field tangential component, while boundary
conditions across the graphene (at interface $z=0$) can be expressed
in matrix form
\begin{eqnarray}
\left(\begin{array}{c}
h_{y||k_{x}}^{(b)}\left(0\right)\\
e_{x||k_{x}}^{(b)}\left(0\right)
\end{array}\right)=\hat{Q}^{(g)}\left(\begin{array}{c}
h_{y||k_{x}}^{(t)}\left(0\right)\\
e_{x||k_{x}}^{(t)}\left(0\right)
\end{array}\right),\label{eq:bc-graphene}\\
\hat{Q}^{(g)}=\left(\begin{array}{cc}
1 & -\frac{4\pi}{c}\overline{\sigma^{(g)}\left(\omega\right)}\\
0 & 1
\end{array}\right),\nonumber 
\end{eqnarray}
and average graphene's conductivity $\overline{\sigma^{(g)}\left(\omega\right)}$
will be different for the acoustic and graphene modes. 

\begin{figure}
	\includegraphics[width=170mm]{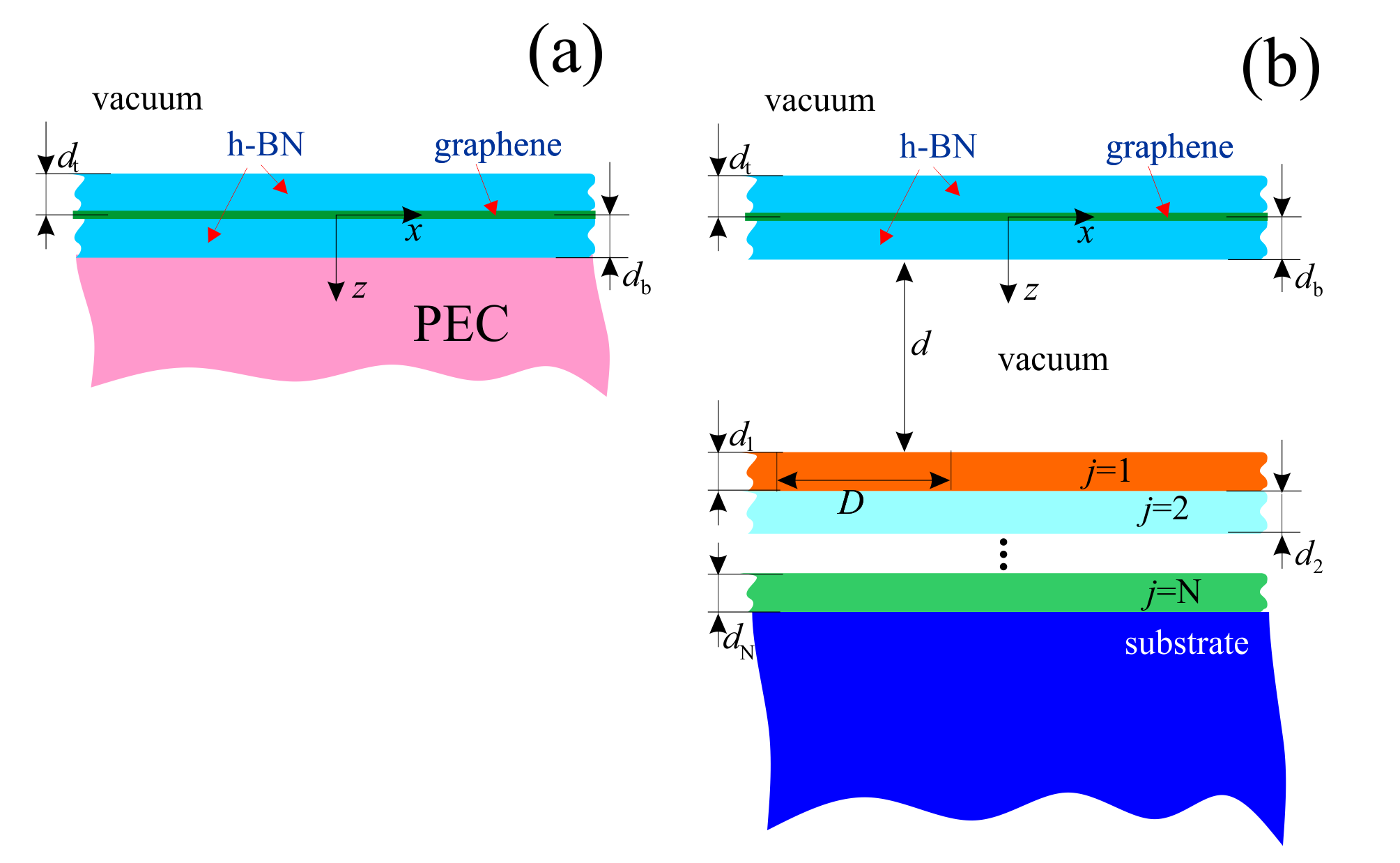}
	
	\caption{Equivalent schemes for the dispersion relation of acoustic {[}panel
		(a){]} and graphene modes {[}panel (b){]}.}
	
	\label{fig:equivalent-schemes}
\end{figure}
For the acoustic mode we use the equivalent scheme, depicted in Fig.\,\ref{fig:equivalent-schemes}(a)
-- a graphene layer, cladded between two hBN layers, and deposited
on top of PEC. This structure is equivalent to one, depicted in Fig.\,\ref{fig:geometry},
but when $W=0$. Notice, that layers at $z>d+d_{b}$ are screened
by PEC. The dispersion relation can be obtained by consequent multiplication
of matrices in Eqs.\,\ref{eq:HE-vacuum-kx}, \ref{eq:HE-top-hBN-kx},
\ref{eq:bc-graphene}, and \ref{eq:HE-bottom-hBN-kx}, thus obtaining
expression for the electromagnetic fields on the surface of PEC as
\begin{eqnarray}
\left(\begin{array}{c}
h_{y||k_{x}}^{(b)}\left(d_{b}\right)\\
e_{x||k_{x}}^{(b)}\left(d_{b}\right)
\end{array}\right)=\hat{F}_{k_{x}}^{(v,tot)}\left(\begin{array}{c}
0\\
H_{y||k_{x}}^{(r)}
\end{array}\right),\label{eq:heb}
\end{eqnarray}
where $\hat{F}_{k_{x}}^{(v,tot)}=\left[\hat{Q}_{k_{x}}^{(b)}\left(d_{b}\right)\right]\hat{Q}^{(g)}\left[\hat{Q}_{k_{x}}^{(t)}\left(d_{t}\right)\right]\hat{F}_{k_{x}}^{(v)}$
is the total field matrix {[}similar to one in Eq.\,\ref{eq:Fvtot}{]}.
Here the expression for the graphene's average conductivity in boundary
condition matrix \ref{eq:bc-graphene} is defined as 
\begin{eqnarray*}
	\overline{\sigma^{(g)}\left(\omega\right)}=\frac{1}{W}\intop_{-W/2}^{W/2}\sigma^{(g)}\left(\omega,x\right)dx,
\end{eqnarray*}
i.e. averaged in the region above single PEC bar. The boundary condition
on the PEC is zero tangential component of electric field on PEC's
surface, i.e. $e_{x||k_{x}}^{(b)}\left(d_{b}\right)=0$. Being substituted
into \ref{eq:heb}, it gives the dispersion relation for the acoustuc
mode in the form
\begin{eqnarray*}
	\left[\hat{F}_{k_{x}}^{(v,tot)}\right]_{22}=0.
\end{eqnarray*}

To obtain the dispersion relation of graphene modes the equivalent
scheme can be obtained from that in Fig.\,\ref{fig:geometry} by
putting $W=D$. In this case the periodic structure is substituted
by vacuum of finite thickness $d$, as it is shown in Fig.\,\ref{fig:equivalent-schemes}(b).
The fields at the finite vacuum bottom boundary $z=d_{b}+d$ can be
obtained by consecutive multiplication of matrices in Eqs.\,\ref{eq:HEs-substrate-kx}
and \ref{eq:HE-j-kx}, thus giving
\begin{eqnarray}
\left(\begin{array}{c}
h_{y||k_{x}}^{(1)}\left(d_{b}+d\right)\\
e_{x||k_{x}}^{(1)}\left(d_{b}+d\right)
\end{array}\right)=\hat{F}_{k_{x}}^{(s,tot)}\left(\begin{array}{c}
H_{y||k_{x}}^{(s)}\\
0
\end{array}\right),\label{eq:he1}
\end{eqnarray}
where $\hat{F}_{k_{x}}^{(s,tot)}=\left[\prod_{j=1}^{N}\hat{Q}_{k_{x}}^{(j)}\left(-d_{j}\right)\right]\hat{F}_{k_{x}}^{(s)}$
is the total field matrix, whose structure is similar to that in Eq.\,\ref{eq:Fstot}.
At the same time, fields at both sides of finite thickness vacuum
can be related via transfer matrix as
\begin{eqnarray}
\left(\begin{array}{c}
h_{y||k_{x}}^{(1)}\left(d_{b}+d\right)\\
e_{x||k_{x}}^{(1)}\left(d_{b}+d\right)
\end{array}\right)=\hat{Q}_{k_{x}}^{(v)}\left(d\right)\left(\begin{array}{c}
h_{y||k_{x}}^{(b)}\left(d_{b}\right)\\
e_{x||k_{x}}^{(b)}\left(d_{b}\right)
\end{array}\right),\label{eq:HE-vac-kx}
\end{eqnarray}
where $\varepsilon_{xx}^{(v)}\left(\omega\right)=\varepsilon_{zz}^{(v)}\left(\omega\right)=1$.
Combining Eqs.\,\ref{eq:heb}, \ref{eq:he1}, and \ref{eq:HE-vac-kx},
we have 
\begin{eqnarray*}
	\hat{F}_{k_{x}}^{(s,tot)}\left(\begin{array}{c}
		H_{y||k_{x}}^{(s)}\\
		0
	\end{array}\right)=\hat{Q}_{k_{x}}^{(v)}\left(d\right)\hat{F}_{k_{x}}^{(v,tot)}\left(\begin{array}{c}
		0\\
		H_{y||k_{x}}^{(r)}
	\end{array}\right),
\end{eqnarray*}
from which the dispersion relation for graphene mode can be obtained
in the form
\begin{eqnarray}
\left[\left(\hat{F}_{k_{x}}^{(s,tot)}\right)^{-1}\hat{Q}_{k_{x}}^{(v)}\left(d\right)\hat{F}_{k_{x}}^{(v,tot)}\right]_{22}=0.\label{eq:dr-graphene}
\end{eqnarray}
Notice, that the averaging of graphene's conductivity in boundary
condition matrix \ref{eq:bc-graphene} {[}which in its turn is included
into matrix $\hat{F}_{k_{x}}^{(v,tot)}$ in dispersion relation \ref{eq:dr-graphene}{]}
is performed over the region above the gap between PEC bars, i.e.
\begin{eqnarray*}
	\overline{\sigma^{(g)}\left(\omega\right)}=\frac{1}{D-W}\intop_{W/2}^{D-W/2}\sigma^{(g)}\left(\omega,x\right)dx.
\end{eqnarray*}


\end{document}